\documentstyle[aps,prl,epsfig]{revtex}

\newcommand{\be}{\begin{equation}}
\newcommand{\ee}{\end{equation}}
\newcommand{\ra}{\rightarrow}
\newcommand{\wi}{\omega_{\infty}}
\newcommand{\ve}{\varepsilon}
\newcommand{\integ}{\int_{0}^{\infty}}
\newcommand{\smfrac}[2]{{\textstyle{#1\over#2}}}
\def\half{\smfrac{1}{2}}
\def\thrhalf{\mbox{\tiny{$3/2$}}}
\def\fhalf{\mbox{\tiny{$5/2$}}}

\begin{document}
\title{Charged Scalar-Tensor Boson Stars: Equilibrium, Stability and
  Evolution}

\author{A. W. Whinnett$^1$\thanks{Email: A.W.Whinnett@qmw.ac.uk}
\& Diego F. Torres$^2$\thanks{Email: dtorres@venus.fisica.unlp.edu.ar}}

\address{$^1$ School of Mathematical Sciences, Queen Mary and Westfield
College, University of London, United Kingdom\\
$^2$ Departamento de F\'{\i}sica, Universidad Nacional de La Plata,
C.C. 67, 1900 La Plata, Buenos Aires, Argentina}

\maketitle

\begin{abstract}

We study charged boson stars in scalar-tensor (ST) gravitational
theories. We analyse the weak field limit of the solutions and 
analytically show that there is a maximum charge to mass ratio for the
bosons above which the weak field solutions are not stable. This
charge limit can be greater than the GR limit for a wide class of ST
theories. We numerically investigate strong field solutions in 
both the Brans Dicke and power law ST theories. We find that the 
charge limit decreases with increasing central boson density. 
We discuss the gravitational evolution of charged and 
uncharged boson stars in a cosmological
setting and show how, at any point in its evolution, the physical
properties of the star may be calculated by a rescaling of a solution
whose asymptotic value of the scalar field is equal to its initial
asymptotic value. We focus on evolution in which the particle number
of the star is conserved and we find that the energy and central
density of the star decreases as the cosmological time increases.
We also analyse the appearance of the scalarization
phenomenon recently discovered for neutron stars configurations and, finally,
we give a short discussion on how making the correct choice of mass
influences the argument over which conformal frame, the Einstein frame
or the Jordan frame, is physical.

\end{abstract}


\section{Introduction}\label{INT}

Boson stars are localised, asymptotically flat
configurations of gravitationally bound zero
temperature bosons. Mathematically, the boson field is described by a
complex wave function $\psi$ whose Lagrangian possesses an internal
$U(1)$ symmetry that gives
rise to a conserved charge $N$, interpreted as the total number of
bosons. The time dependence of $\psi$ does not appear in the field
equations, and the solutions describe bound eigenstates of $\psi$
with a number of nodes that increases with the energy. 
The first boson star solutions were found by Kaup \cite{Kaup1} and
independently by Ruffini \& Bonazzola \cite{RB}, who studied
spherically symmetric stars in General Relativity (GR). They
found that the solutions were qualitatively similar to those describing
neutron stars and white dwarfs, although they were of much
smaller mass. Colpi, Shapiro \& Wasserman \cite{Colpi} 
extended their work by examining GR
boson stars whose matter Lagrangian includes a quartic 
self-interaction term. This additional term contributes to the pressure of
the star, increasing both its mass and particle number.
More general self-interaction terms can be considered. The first
research on this subject was undertaken by Lee and co-workers
\cite{FLP} who, in a series of papers, studied the properties of
non-topological solitons. These are boson stars whose self-interaction
term allows localised, non-singular solutions to exist even in the
absence of gravity. To accomplish this, the self-interaction must have
attractive terms and must be at least sixth order in the boson field 
amplitude.
The work of Lee et al.\  was performed independently of the main body
of boson star research. For the case of non-soliton stars, it is not 
immediately obvious how the inclusion of a non-quartic
self-interaction affects the properties of the star. 
This problem will be addressed elsewhere \cite{ST};
for our purposes, it will be sufficient to only consider the usual quartic
self-interaction term. For reviews see Ref.\  \cite{reviews}. 

Charged boson stars in GR were first studied by Jetzer \& van der Bij 
\cite{JVB} who found that the inclusion of the charge increases the star
mass and particle number, in much the same way as the inclusion of the
quartic self-interaction. In addition, they found that there was an
upper limit on the charge to mass ratio of the bosons above which no 
non-singular solutions could be found: 
stars made up of bosons whose charge is
beyond this limit generate a repulsive Coulomb field that overcomes
the gravitational attraction, regardless of the number of bosons that
make up the star. The stability properties of charged boson stars in GR
were studied by Jetzer \cite{NEW-JETZER} before the advent of the use
of catastrophe theory in stellar equilibrium. Also, Jetzer and coworkers
\cite{JEZ} have discussed the influence of charged boson stars on the 
stability of the vacuum and have considered their contribution to
cosmological dark matter.

Recently, there has been renewed interest in alternative theories of
gravity, in particular scalar-tensor (ST) theories, in which the usual
metric gravitational field is augmented by a scalar field $\phi$ which
couples to the curvature via a parameter $\omega(\phi)$. The strength
of this coupling increases with $1/\omega$. The simplest
of these is Brans Dicke (BD) theory \cite{BD} in which
$\omega$ is constant. These present studies of ST theories are motivated by
the fact that they appear as the low energy limit of string theory
\cite{STRINGS} (the simplest string effective action being the
$\omega=-1$ BD action) and that scalar gravitational fields arise from
dimensional reduction of higher dimensional theories (see
Ref.\  \cite{ZHUK} and citations therein). These theories may better
describe gravity in the early Universe and, in addition, several models
of inflation are driven by the same scalar field of ST gravity \cite{INF}.
Finally, ST theories provide a self-consistent framework in which one
can study the possible variation of the gravitational coupling $G$.

The strongest constraints on ST theories are usually assumed to
come from the solar system weak field tests \cite{Will}. As is well known,
observations constrain the BD parameter to have a value of $\omega>500$
to within $1\sigma$. However, more general ST theories can satisfy the
rather stringent constraints determined at the current epoch, while still
differing considerably from GR in the past \cite{BARROW-PRD}. In
addition, it has been shown that strong non-perturbative ST
gravitational effects may arise in ST neutron star
configurations at the present epoch, even when the parameters of the
theory accord well with observational tests \cite{SCALARIZATION}.

The current popularity of alternative models of gravity has
prompted several authors to study boson stars in the framework of ST
theory. The first of these 
were Gunderson \& Jensen \cite{Gund}, who considered
BD boson stars with $\omega=6$. More general ST boson stars
were considered by Torres \cite{Torres1} and later by Comer \&
Shinkai \cite{CS}. These studies showed that
the inclusion of the $\phi$ field introduces no qualitative change in
the solutions, although in all cases, boson stars have 
a slightly smaller mass than their GR counterparts. 

Spherically symmetric charge distributions in ST theory
have also been investigated before. Singh \& Usham
\cite{SINGH} studied particular solutions of the BD field equations
with a charged perfect fluid source in Dicke's conformally 
transformed units. van der Bergh \cite{BERGH} analysed the case of a ST
spacetime containing an electrostatic field, with traceless 
energy-momentum tensor, and found both naked time-like singularities and 
non-singular solutions. Reddy \cite{REDDY} and Reddy \& Rao \cite{RR}
also obtained spherically symmetric, static, conformally flat solutions
of the BD electro-vacuum field equations. Neither of these authors,
however, have considered the case of massive particles bounded 
to form a star, and the existing literature does not provide an analysis
of the charged boson star system as we do here. 

While charged stars, either in GR or in a ST theory, are interesting
objects from a theoretical standpoint, until last year
it was thought unlikely
that they would play a significant astrophysical role. The argument to
support this view was that of selective accretion:
any charged object will naturally accrete matter
of opposite sign and ultimately become neutral, as would, for
example, a Kerr-Newmann black hole. However, a recent study
by Punsly \cite{PUNSLY} established that a charged rotating
black hole will be able not only to preserve its charge, 
but will also generate observable gamma rays. 
In an extremely simplified view of his
model, he considered the black hole to be
surrounded by a ring of opposite
charge, but otherwise isolated in interstellar space. The ring,
while neutralizing the whole system as seen by an observer at spatial
infinity, would allow the black hole to keep its charge. Magnetic flux lines
would enter the hole through its axis of symmetry and pair creation would
produce gamma rays.
This process may be important in understanding the nature of some of the
unidentified galactic gamma ray sources recently discovered by the 
NASA Compton satellite, in its 
EGRET experiment \cite{EGRET}.

The already initiated program of searching for observational signals 
of boson stars has lead us to some significant analogies with black holes
\cite{LIDDLESCHUNCK}. Therefore, it is natural to expect that research into
the subject of charged boson stars would extend this analogy. For
instance, it may be possible that 
a rotating charged boson star may replace the central engine of Punsly's
mechanism, although a careful analysis should be made to prove or
disprove this statement (something which is far from the purpose
of the present work). It is worth noting, however, that
the more realistic models of (uncharged) rotating boson stars have been
investigated and solutions were shown to exist \cite{ROT}.
There is no reason to suppose that rotating charged boson star solutions 
do not also exist. All in all, it is encouraging that charged objects 
are found to produce testable effects from which their
astrophysical signatures may be extracted: the presence of charge
allows a wider range of phenomena to be observable. In the context of our
present work, boson stars, either charged or 
uncharged, are relativistic constructs that may help us understand the 
influence of a different gravitational theory and its cosmology on the
astrophysical world.

In a ST theory, at the level of the action, Newton's gravitational 
constant $G$ is replaced by a field $G^{\star}=G/\phi$. 
In a cosmological setting $G^{\star}$ will evolve
with time, giving rise to the twin phenomena of gravitational memory
and gravitational evolution \cite{GRAV-MEM}. 
Consider a compact object in equilibrium that
forms at some time $t_{1}$ at which the gravitational coupling
strength is $G^{\star}_{1}$. The value of $G^{\star}_{1}$ will determine to
some extent the structure of the object. At some later 
cosmological time, $G^{\star}$
will have evolved to some new value $G^{\star}_{2}$ and the object 
may either ``remember'' its local value of $G_{1}^{\star}$ or evolve to
some new configuration in which its local value of $G^{\star}$ matches the
cosmological value $G^{\star}_{2}$. 
In actual fact, one would expect the behaviour of
the object to lie somewhere between these two extremes. Boson
stars seem to be ideal candidates for the study of these phenomena:
they are relatively simple objects whose equilibrium solutions are
easy to find, they are non-singular and they are fully relativistic.
Gravitational memory and gravitational evolution in boson stars were 
first analysed by Torres, Liddle \& Schunck \cite{TLS}, who constructed
sequences of static configurations at different cosmological
times. One can also consider the evolution of other compact objects, such as
the white dwarfs studied by Garc\'{\i}a-Berro et al.\  \cite{BERRO} and
Benvenuto et al.\  \cite{BENVE},
who found that the variation of $G^{\star}$ produced observable
changes in the stars' luminosities. 

The bosons that make up a boson 
star are assumed to have identical mass $m$
and, as we shall show in detail in the following Section, 
$m$ and the
Planck mass $M_{pl}$ serve as a scale against which all physical
properties of the star may be measured. Here we shall briefly outline
the range of physical magnitudes possible for given choices of
$m$. For a mini-soliton star (made of uncharged non self-interacting
bosons), the choice $m=30$ GeV implies that the mass of the star is
of order $10^{10}$ kg, which is 20 orders of magnitude smaller that the
solar mass $M_{\odot}$, while its radius is of order $10^{-17}$ m. 
This gives a density of around $10^{48}$ times that of a neutron star
\cite{reviews} and these objects are obviously highly
relativistic. Only when the boson mass is reduced to around
$10^{-10}$eV does the star resemble a more conventional stellar
object. As we have mentioned above, including a self-interaction
potential $V(\psi)$ can substantially alter these figures. The
importance of the self-interaction term
is measured by the ratio $V(\psi)/(m^{2}|\psi|^{2})$,
which for the quartic self-interaction considered by Colpi et al.\
\cite{Colpi} is approximately $\lambda M_{pl}^{2}/m^{2}$, where $\lambda$
is a constant and one assumes that $|\psi|\sim
M_{pl}$. In this case, the self-interaction can only be neglected if
$\lambda$ is exceedingly small:
$\lambda\ll m^{2}/M_{pl}^{2}=6.7\times
10^{-39}$ GeV$^{-2}$m$^{2}$. Rewriting the potential as the
dimensionless number $\Lambda:=\lambda M_{pl}^{2}/(4\pi m^{2})$, the
equilibrium solutions can be parameterised by $\Lambda$ and for
$\Lambda\gg 1$, the boson star masses scale as 
$\Lambda^{1/2} M_{pl}^2/m \sim
\lambda^{1/2} M_{pl}^3 / m^2 \sim \lambda^{1/2} M_{ch}$, where
$M_{ch}$ is the Chandrasekhar mass (approximately $1 M_\odot$). 
For instance, if $m = 1$ GeV, the star has a mass of $1.6
\sqrt{\lambda}M_\odot$ and a radius of $6\sqrt{\lambda}$km. 
If, instead, we take $m=1$ MeV, then the
mass becomes $10^{6}\sqrt \lambda M_{\odot}$ and the radius becomes
$10^6 \sqrt \lambda$ km, which is similar to the
Sun's radius but encompasses one million solar masses. For
non-topological soliton stars, the mass was found to scale as
$M_{pl}^{4}/m^{3}$ and one can construct soliton stars of very large
mass and radius (the latter quantity being of the order of a few light
years). The fact that
both $m$ and $\lambda$ are free parameters allows one to construct
boson star solutions with physical properties whose values range over
all astrophysically interesting magnitudes. With the inclusion of the
gauge charge, the range of physical magnitudes will depend upon a third
independent parameter. We shall discuss this in more detail when we
analyse numerical solutions in Section \ref{NS}.

The rest of this work is organized as follows. Section \ref{FORM}
introduces the basic formalism, both for gravity and the stellar model. 
An analysis of the charge, the mass definition and the asymptotic limit 
is made in Section \ref{CM}. In Section \ref{BV} we discuss the boundary 
values for all fields and provide scaling relations valid for our field 
equations, which we shall later use. We then analytically prove, in
Section \ref{WF},
that there exist a maximum charge for weak field configurations
(characterised by small central densities) which differs from the GR result 
(we shall also explicitly derive this last result
as the correct limit of the ST case).
In Section \ref{NS} we numerically analyse strong field solutions for
both BD gravity and a representative power law ST theory. 
These latter solutions
demonstrate how ST boson stars may produce observable ST effects at
the current epoch.
Section \ref{GE} is devoted to the
discussion of gravitational memory and evolution of boson stars, both charged
and uncharged. We give our final comments in Section \ref{CONCLUSIONS}.

\section{Formalism}\label{FORM}

The system we study is that of a complex scalar field $\psi$
with a $U(1)$ charge in the framework of ST
gravity. Its action is given by

\be\label{action}
   S=\frac{1}{16\pi }\int d^{4}x\sqrt{-g}\left[ \phi R-\frac{\omega(\phi )}
   \phi g^{\mu \nu }\phi _{,\mu }\phi _{,\nu }+ 16\pi L_m\right],
\ee
where $g=$ Det$(g_{\mu\nu})$, $R$ is the scalar curvature, $\omega$ is the
coupling function and $L_m$ is the matter Lagrangian. We take 
$L_m$ to be
\be\label{Lmatter}
   L_m=-\frac{1}{2} g^{\mu\nu} ({\overline D}_{\mu} {\overline \psi})
   (D_{\nu} \psi)-\frac{1}{2} m^2 \psi{\overline\psi} -\frac{1}{4} \lambda 
   (\psi{\overline\psi})^2
   -\frac{1}{16\pi}F^{\mu\nu}F_{\mu\nu}
\ee
where
\be
F_{\mu\nu}=\partial_{\mu}{\cal A}_\nu-\partial_{\nu}{\cal A}_{\mu},
\ee
\be
D_{\mu}=\partial_{\mu}+ie{\cal A}_\mu,
\ee
${\cal A}_{\mu}$ is the gauge field and the over-bar denotes complex
conjugation. This Lagrangian is invariant under a gauge transformation
of the $U(1)$ group. This implies the existence of a conserved current
\be\label{current}
   {\cal J}^\mu=\sqrt{-g} g^{\mu\nu} \left[ ie \left(
   {\overline\psi}\partial_\nu \psi -\psi \partial_\nu{\overline\psi} \right) -
   2e^2 {\cal A}_\nu \psi{\overline\psi} \right],\;\;\;\;\;\;\;
   \partial_{\mu}{\cal J}^{\mu}=0.
\ee

Varying the action with respect to $g^{\mu\nu}$ and
$\phi$ we obtain the field equations
\be\label{field1}
   R_{\mu\nu}-\frac{1}{2} g_{\mu\nu}R=\frac{8\pi}{\phi}T_{\mu\nu}
   +\frac{\omega(\phi)}{\phi^2} \left[ \phi_{,\mu} \phi_{,\nu}-
   \frac{1}{2} g_{\mu \nu }\phi^{,\alpha}\phi_{,\alpha}\right]+ 
   \frac{1}{\phi} \left[ \phi_{,\mu;\nu} -g_{\mu\nu} \Box\phi \right]
\ee
and
\be\label{field2}
   \Box \phi =\frac{1}{2\omega+3} \left[ 8\pi T-\frac{d\omega}{d\phi}
   \phi^{,\alpha}\phi_{,\alpha}\right],
\ee
where
\be
   T_{\mu\nu}=({\overline D}_{\mu}{\overline\psi}) (D_{\nu}\psi)
   +\frac{1}{4\pi}F_{\mu}^{\;\;\sigma}
   F_{\nu\sigma}-\half g_{\mu\nu}\left[
   g^{\alpha\beta}({\overline D}_{\alpha}{\overline\psi})(D_{\beta}\psi)
   +m^2\psi{\overline\psi}+\half\lambda(\psi{\overline\psi})^2+\frac{1}{8\pi}
   F_{\alpha\beta}F^{\alpha\beta}\right]
\ee
is the energy-momentum tensor for the matter fields and $T$ is its trace.

Varying the action with respect to the mater fields $\psi$,
${\overline\psi}$ and ${\cal A}_{\mu}$ we obtain the boson wave equations
\be\label{field3}
   g^{\alpha\beta}\nabla_{\alpha}D_{\beta}\psi=
   (m^2+\lambda{\overline\psi}\psi)\psi-ieg^{\alpha\beta}{\cal
   A}_{\alpha}D_{\beta}\psi,\;\;\;\;
   g^{\alpha\beta}\nabla_{\alpha}{\overline D}_{\beta}{\overline\psi}=
   (m^2+\lambda{\overline\psi}\psi){\overline\psi}+ieg^{\alpha\beta}{\cal
   A}_{\alpha}{\overline D}_{\beta}{\overline\psi}
\ee
and the Maxwell equation
\be\label{field4}
   \nabla_{\nu}F^{\mu\nu}=2\pi J^{\mu},
\ee
where
\be
   J^{\mu}:=\frac{1}{\sqrt{-g}}{\cal J}^{\mu}.
\ee

We consider static, spherically symmetric solutions and write the line
element in the Schwarzschild form

\be \label{metric}
   ds^2=-B(r) dt^2 + A(r) dr^2 +r^2 d\Omega^2,
\ee
where $d\Omega^{2}$ is the line element of the unit 2-sphere. We look
for solutions of minimum energy. It has been shown \cite{FLP} that for
spherically symmetric systems of uncharged bosons, since
$L_{m}$ is universally coupled to the curvature, the form of
$\psi$ compatible with this requirement is
\be\label{psidef}
   \psi(r,t)=\chi(r) \exp{[i \varpi t]},
\ee
where $\varpi$ is a real positive constant. To prove this, one
minimises the total energy of the boson star subject to the constraint
that the particle number (which we shall define below) be
conserved. The parameter $\varpi$ plays the role of a Lagrange
multiplier and this result is independent of the form of the boson
self-interaction $V(\phi)$. This result also holds when one includes a
$U(1)$ charge \cite{AW2}, and so we take Eq. \ref{psidef} to be valid
in the present work.

Equation (\ref{psidef}) is also 
consistent with the
assumption that $g_{\mu\nu}$ is static. The scalar field $\phi$ inherits
the symmetries of the line element and is a function of $r$, while the
gauge field may be chosen such that we have only electric charges
and no magnetic ones, so we set
\be
\label{gauge}
 {\cal A}_\mu=({\cal A}_{0}(r),0,0,0).
\ee

Before we explicitly write down the full set of equations of structure
of the star, we shall introduce some dimensionless quantities  
which will simplify the comparison with the previous work cited in
Section \ref{INT}. Using the boson mass $m$ and the
present value of the Planck mass $M_{pl}:=\sqrt{1/G}$ to define 
length and energy scales, the new quantities are:

\be
\label{Omega}
   \Omega(r)=\frac{\varpi+e{\cal A}_{0}}{m},
\ee
\be
\label{sigma}
   \sigma=\sqrt{4\pi} \frac{\chi(r)}{M_{pl}}
\ee
and
\be\label{L}
   \Lambda=\frac{\lambda}{4\pi} \left( \frac{M_{pl}}{m} \right)^2
\ee
for the boson sector,

\be\label{C}
   C=\frac{{\cal A}_0}{M_{pl}}
\ee
and
\be\label{q}
   q=\frac{eM_{pl}}{m}
\ee
for the gauge field, and

\be\label{Phi}
   \Phi=\frac{\phi}{M_{pl}^2}
\ee
for the Brans-Dicke field. We also introduce a rescaled radial
coordinate $x$ defined by
\be\label{x}
   x:=mr.
\ee
In terms of these quantities, the independent components of the field 
equations (\ref{field1}), (\ref{field2}), (\ref{field3}) and 
(\ref{field4}) are

\begin{itemize}

\item The generalized Einstein equations:

\begin{eqnarray}\label{dAdx}
   A^{\prime}=Ax\left[
   \frac{\sigma^{\prime 2}(2\omega+1)}{\Phi(2\omega+3)}
   +\frac{\sigma^{2}A}{\Phi(2\omega+3)}\left(\frac{\Omega^{2}}{B}
   (2\omega+5)+2\omega-1\right)  
   +\frac{\sigma^4 A \Lambda (2\omega-1)}{2\Phi(2\omega+3)}
   \right. \nonumber \\ \left.
   -\frac{B^{\prime}\Phi^{\prime}}{2B\Phi}
   +\frac{\omega\Phi^{\prime 2}}{2\Phi^{2}}
   -\frac{\Phi^{\prime 2}}{\Phi(2\omega+3)}\left(\frac{d\omega}{d\Phi}\right)
   +\frac{C^{\prime 2}}{B\Phi}+\frac{1}{x^2}(1-A)\right]
\end{eqnarray}
\be\label{dBdx}
   B^{\prime}=\frac{x}{2\Phi+x\Phi^{\prime}}\left[2B\sigma^{\prime 2}
   +2\sigma^{2}A(\Omega^{2}-B)-\sigma^4 A B \Lambda
   -\frac{4B}{x}\Phi^{\prime}+\frac{B\omega}{\Phi}\Phi^{\prime 2} 
   -2C^{\prime 2}+\frac{2B\Phi}{x^{2}}(A-1)\right]
\ee

\item The scalar field equation:

\be\label{ddPdxx}
   \Phi^{\prime \prime} = \Phi^{\prime} \left( \frac{A^\prime}{2A} -
   \frac{B^\prime}{2B} - \frac{2}{x} \right) + \frac{2A}{2\omega+3} 
   \left[ \left(
   \frac{\Omega^2}{B}-2 \right)\sigma^2 -\frac{\sigma^{\prime 2}}{A} - 
   \Lambda \sigma^4 \right] -
   \frac{1}{2\omega+3}
   \frac{d\omega}{d\Phi} 
   \Phi^{\prime 2}
\ee

\item The boson field equation:

\be\label{ddsdxx}
   \sigma^{\prime \prime} = \sigma^{\prime} \left( \frac{A^\prime}{2A} -
   \frac{B^\prime}{2B} - \frac{2}{x} \right) - A \left[ \left(
   \frac{\Omega^2}{B}-1 \right)\sigma - \Lambda \sigma^3 \right]
\ee

\item The Maxwell equation:

\be\label{ddCdxx}
   C^{\prime \prime} =  C^{\prime} \left( \frac{B^\prime}{2B} +
   \frac{A^\prime}{2A} - \frac{2}{x} \right) + qA \Omega \sigma^2
\ee

\end{itemize}
Here a prime denotes $d/dx$. We mainly consider two forms of $\omega$:
the Brans Dicke coupling in which $\omega$ is constant and a power law
coupling, where $2\omega+3=4n/3\Phi^{n}$. 

Equations (\ref{dAdx}) to (\ref{ddCdxx}) reduce to the general
relativistic ones \cite{JVB} when $\omega \ra \infty$, 
$\Phi\ra 1$ and $d\omega/d\Phi\ra 0$. Note, however, that this is not 
a simple issue
whenever the trace of the matter energy-momentum tensor vanishes 
\cite{FARAONI-W}. This is not the case here, although we still 
verify the general relativistic limit for large
values of $\omega$.

\section{Charge, Mass and the Asymptotic Limit}\label{CM}

We shall first look for solutions that are asymptotically flat
in the sense that, in the limit $x\ra\infty$,
\be\label{asymflat}
   A\sim 1+{\cal O}\left(\frac{1}{x^{n}}\right),\;\;\; 
   B\sim B_{\infty}+{\cal O}\left(\frac{1}{x^{n}}\right),\;\;\; 
   \Phi\sim\Phi_{\infty}+{\cal O}\left(\frac{1}{x^{n}}\right),\;\;\; 
   C\sim C_{\infty}+{\cal O}\left(\frac{1}{x^{n}}\right),
\ee 
where $n\ge 1$ and the subscript
`$\infty$' denotes values at space-like infinity. We also require 
$\sigma$ to vanish at least as fast as $x^{-1}$ so that the boson
matter is localised and the solutions describe bound eigenstates of
$\psi$. In fact, as we shall show
below, the conditions above imply that $\sigma$ falls off
exponentially as $x\ra\infty$. In addition, the field
equations (\ref{dAdx}) to (\ref{ddCdxx}) are invariant under the rescaling
$C\ra C+c$, $\varpi\ra\varpi-qc$ where $c$ is a
constant, and we use this remaining gauge freedom to set
\be\label{asymC}
   C_{\infty}=0.
\ee
This implies that $\Omega_{\infty}=\varpi/m$.

Given the asymptotic flatness conditions, we may define  
expressions for the total charge and mass 
(or energy) of the boson stars. From 
the conserved current equation (\ref{current}) one may derive a
conserved (time independent) charge
\be\label{intQ}
   Q=q\integ\Omega\sqrt{\frac{A}{B}}\;\sigma^2 x^2\;dx.
\ee
This quantity may be interpreted as the star's total electrical charge per
unit boson mass. Since all of the bosons are assumed to carry identical
charge per unit mass $q$, we may also define a conserved particle
number
\be\label{defN}
   N:=\frac{Q}{q}=\integ\Omega\sqrt{\frac{A}{B}}\;\sigma^2 x^2\;dx.
\ee
Due to the fact that 
we are using the boson mass $m$ as the mass scale, the numerical
value of $N$ is also equal to the value of
the `rest mass' of the star, a term we use to describe the sum of
the masses of the bosons measured in some non-gravitational way. Then
the particle number is measured in units of $M_{pl}^{2}/m^{2}$ while
the rest mass is measured in units of $M_{pl}^{2}/m$. We
also define a Newtonian mass
\be\label{defMN}
   M_{N}:=\frac{N}{\Phi_{\infty}}
\ee
which measures the active gravitational mass of the star whose bosons are
dispersed to infinity. 

Integrating the Maxwell equation (\ref{ddCdxx}) we find an alternative
expression for the total charge:
\be\label{Qlimit}
   Q=\integ \frac{d\;}{dx}\left(\frac{x^2}{\sqrt{AB}}
   \frac{dC}{dx}\right)\;dx=\lim_{x\to\infty}
   \left(x^2 \frac{dC}{dx}\right),
\ee
where we have use the asymptotic flatness conditions to derive the
second equation. In an analogous way, we define the `scalar charge'
\be\label{Scharge}
   S:=\lim_{x\to\infty}
   \left(x^2 \frac{d\Phi}{dx}\right).
\ee
Note that this is not in general a conserved quantity as there is
no internal
symmetry associated with $\Phi$ in the action (\ref{action}). We
introduce it here since it is included in the expression for the 
mass of the star which we shall give below. Integrating the wave equation
(\ref{ddPdxx}) gives an alternative expression for the scalar charge:
\be\label{intS}
   S= \integ x^2 \left[\Phi^{\prime} \left( \frac{A^\prime}{2A} -
   \frac{B^\prime}{2B} \right) + \frac{2A}{2\omega+3} 
   \left[ \left(
   \frac{\Omega^2}{B}-2 \right)\sigma^2 -\frac{\sigma^{\prime 2}}{A} - 
   \Lambda \sigma^4 \right] - \frac{1}{2\omega+3}
   \frac{d\omega}{d\Phi} \Phi^{\prime 2}\right]\;dx.
\ee

The generalisation of the Schwarzschild mass is defined implicitly by
the familiar relation
\be\label{defMS}
   A:=\left(1-\frac{2M(x)}{x}\right)^{-1}.
\ee
Note that this equation does not explicitly include the scalar field
$\Phi$: here the gravitational coupling strength $G^{*}=G/\Phi$ has
been factored into $M$. Using Eq.\  (\ref{dAdx}) the Schwarzschild 
mass may be written as the integral
\begin{eqnarray}\label{intM}
   M(x)=\int_{0}^{x}d{\tilde x}\;\frac{{\tilde x}^2}{2}
   \left[\frac{\omega\Phi^{\prime 2}}{2A\Phi^{2}}
   -\frac{\Phi^{\prime 2}}{A\Phi(2\omega+3)}\left(\frac{d\omega}{d\Phi}\right)
   +\frac{\sigma^{2}}{\Phi(2\omega+3)}\left(\frac{\Omega^{2}}{B}
   (2\omega+5)+2\omega-1\right)\right.  \nonumber \\ \left.
   +\frac{\sigma^{\prime 2}(2\omega+1)}{A\Phi(2\omega+3)}
   +\frac{\sigma^4 \Lambda (2\omega-1)}{2\Phi(2\omega+3)}
   -\frac{B^{\prime}\Phi^{\prime}}{2AB\Phi}
   +\frac{C^{\prime 2}}{AB\Phi}\right].
\end{eqnarray}
One can show that 
$M_{\infty}:=\lim_{x\to\infty}M=M_{ADM}$, where $M_{ADM}$ is the ADM
mass of the star. Note that the integrand in Eq.\   (\ref{intM}) contains
non-positive definite terms, so that $M^{\prime}$ and $M$ themselves
may be negative for weak field solutions. This is because the scalar
field terms on the right hand side of Eq.\  (\ref{field1}) may cause
$G_{\mu\nu}$ to violate the dominant energy condition. This fact has
been exploited in the construction of ST wormhole solutions for which
the matter energy-momentum tensor satisfies all of the energy conditions
\cite{ANCHORDOQUI} and even when there is no matter tensor at all \cite{GRAV}.

As shown by Lee in a usually uncited paper 
\cite{Lee}, the correct definition of mass 
for an isolated source in BD gravity is the Tensor mass
\be\label{defMT}
   M_{T}:=M_{ADM}-\frac{S}{2\Phi_{\infty}}.
\ee
This definition is also appropriate for more general ST theories. From 
the Tensor and Newtonian masses, we define the binding energy
\be\label{energy}
   {\cal E}:=M_{T}-M_{N}
\ee
and the fractional binding energy 
\be\label{benergy}
   {\cal B}:=\frac{M_{T}-M_{N}}{M_{N}}
\ee
which measures the binding energy per unit boson mass per boson.
A necessary condition for dynamical stability is that ${\cal E}<0$,
which implies that ${\cal B}<0$.

There are several reasons for preferring $M_{T}$ over $M_{ADM}$ as the
correct definition of the mass of the star in ST gravity.
Lee \cite{Lee} has shown that for a localised source that emits
gravitational radiation, the time rate of change of $M_{T}$ evaluated
at future null infinity is non-positive definite (in other words,
gravitational wave emission can only reduce the Tensor mass of the
source). This is not true for $M_{ADM}$.
Creighton and Mann \cite{CrMann} have given a quasi-local
definition of mass in ST gravity based on a rigorous consideration of
the boundary terms both in the variation of the Lagrangian and in the
formulation of the Hamiltonian. It turns out in either case that the
expression for the mass is given by the usual integral of the
extrinsic curvature on the boundary of the region under consideration,
plus an additional term analogous to the scalar charge defined above.  
For an asymptotically flat spacetime,
and in the appropriate limit, one can show that their mass tends to the Tensor
mass given above. In addition, one can show that, at least for uncharged
boson stars in BD theory, any solution with extremal 
Tensor mass is also one of extremal Newtonian mass (and hence extremal
particle number) \cite{AW}. This result is easily extended to more
general ST theories and is important when one uses catastrophe theory to
determine the stability of a boson star \cite{KMS}: a plot of mass
against particle number for a set of solutions should show ``cusps''
at the extremal values of the particle number. The first of these 
cusps marks the point at which the solutions become dynamically unstable. 

Expanding the metric and matter field equations in powers of $1/x$ about
$x=\infty$ one can write the field equations in a linearised form, similar to
that given in Section \ref{WF}
for the weak field limit. Solving these equations and
using the boundary conditions (\ref{asymflat}) and (\ref{asymC}), we
find that the scalar field and vector potential have the asymptotic
form
\be\label{SClimit}
   \Phi=\Phi_{\infty}-\frac{S}{x}+{\cal O}\left(\frac{1}{x^{2}}\right),
   \;\;\;\;\;\;C=-\frac{Q}{x}+{\cal O}\left(\frac{1}{x^{2}}\right),
\ee
while the line element in this limit is
\be\label{metriclimit}
   ds^2=-\left[1-\frac{2M_{K}}{x}+{\cal O}\left(\frac{1}{x^2}
   \right)\right]dt^2+\left[1+\frac{2M_{ADM}}{x}+{\cal O}\left(
   \frac{1}{x^2}\right)\right]dx^2+x^2 d\Omega^2.
\ee
Here
\be\label{defMK}
   M_{K}:=M_{ADM}-\frac{S}{\Phi_{\infty}}
\ee
is the Keplerian mass of the star. Note that Eq.\   (\ref{SClimit}) can
be obtained directly from Eqs.\   (\ref{Qlimit}) and (\ref{Scharge}).

Substituting the metric components appearing in the asymptotic form of
the line element into the
boson wave equation (\ref{ddsdxx}) gives
\be
   \sigma^{\prime\prime}=-\frac{2\sigma^{\prime}}{x}-\sigma\left(
   1+\frac{2M_{ADM}}{x}\right)\left[\left(\Omega_{\infty}
   -\frac{qQ}{x}\right)^{2}\left(1+\frac{2M_{K}}{x}\right)-1\right]
   +{\cal O}\left(\frac{1}{x^{2}}\right).
\ee
This equation has the solution
\be
   \sigma=x^b e^{-\kappa x}\left[1+{\cal O}\left(\frac{1}{x}\right)
   \right],
\ee
where the constants $\kappa$ and $b$ are given by
\be
   \kappa=\sqrt{1-\Omega_{\infty}^2},\;\;\;\;\;\;
   b=-1+\frac{1}\kappa\left[(2\Omega_{\infty}^2-1)M_{T}
   -\frac{S}{2\Phi_{\infty}}-qQ\Omega_{\infty}\right].
\ee
Hence the boson field falls off exponentially with $x$ in the
asymptotic region. The quantity $\kappa$ may be interpreted as the
reciprocal radius of the star and must be real. This implies that
$\Omega_{\infty}<1$.

We conclude this section with a note on the physical interpretation of
$M_{K}$ and $M_{T}$ and the choice of the physical frame. 
The Keplerian mass is the active gravitational mass measured by a 
non self-gravitating test particle in a circular orbit at space-like
infinity about the star, while the Tensor
mass is the active gravitational mass measured by a `test black hole'
in a similar orbit \cite{Hawk}. By test black hole we mean an object
whose mass is negligible compared to the mass of the boson star and is
made up entirely of gravitational binding energy. Its orbit is
geodesic in the Einstein frame, whose metric ${\widetilde g}_{\mu\nu}$
is related to the physical metric by the conformal transformation
${\widetilde g}_{\mu\nu}=\Phi g_{\mu\nu}$. Calculating the angular
velocity of a circular orbit in the Einstein frame, using Kepler's
third law and transforming the result back into the Jordan 
frame, we find that $M_{T}$ is given by
\be\label{MTorbit}
   M_{T}=\lim_{x\to\infty}\left[\frac
   {r^{2}(\Phi B^{\prime}+B\Phi^{\prime})}{2\Phi+r\phi^{\prime}}\right].
\ee
The expression for $M_{T}$ in the Einstein frame shows that this
quantity is non-negative.

The ADM mass in the Jordan frame cannot be
interpreted in terms of particle orbits. In the limit of 
large $\omega$, $\Phi$ is only
weakly coupled to the curvature and $S$ is small compared with
$M_{ADM}$.  Hence, in this limit the differences between $M_{ADM}$,
$M_{K}$ and $M_{T}$ are negligible and in the exact GR theory all
three definitions are the same.

Recently, Faraoni, Gunzig \& Nardone \cite{FAR-FRAMES} have argued
that the Einstein frame should be considered to be the physical frame,
primarily because the Jordan frame ADM mass is non-positive definite,
even when the matter energy-momentum tensor $T_{\mu\nu}$ 
satisfies all of the energy
conditions. However, we believe that this reasoning is based upon the
wrong choice of mass definition: the physical mass in the Jordan frame
is $M_{T}$ and one can show that $M_{T}$ is positive in the following way.
After a conformal transformation into the Einstein frame, one can show
that the new Einstein tensor will obey the dominant energy condition if
the Jordan frame matter energy-momentum tensor $T_{\mu\nu}$ obeys the
dominant energy condition. Then the Einstein frame ADM mass will be
non-negative, by the usual positive energy theorems. However, by
analysing the effect of a conformal transformation on the definition of
$M_{ADM}$, one can show that the Einstein frame ADM mass is identical
to the Jordan frame Tensor mass, up to a (positive) factor of
$\Phi_{\infty}$ \cite{conframes}.  Recently Santiago \& Silbergleit 
\cite{SANTIAGO} have rewritten the Jordan frame field equations of ST
gravity by defining a new connection that isolates the dynamical degrees of
freedom of the metric gravitational field. Their formalism is
algebraically equivalent to making a conformal transformation into
the Einstein frame, so it not surprising that their result corroborates
with the proof outlined above. The scalar field contribution to the
total energy-momentum of the spacetime can now be made non-negative,
contrary to the previous claim of Faraoni et al. Here we simply use the
conformal relationship between the two frames as a computational aid,
and give our results in the physical Jordan frame.

\section{Boundary Values and Scaling Relations}\label{BV}

In addition to requiring asymptotic flatness, we impose the 
boundary conditions
\be\label{bdry}
   A_{0}=1,\;\;\;\;\sigma_{0}^{\prime}=\Phi_{0}^{\prime}=C_{0}^{\prime}=0,
\ee
where the subscript `0' denotes values at $x=0$. These conditions
ensure that the solutions are regular at the origin and, combined 
with the asymptotic flatness conditions (\ref{asymflat}), they make 
the field equations 
eigenvalue equations for $\varpi$ which automatically lead to the
asymptotic conditions $\sigma_{\infty}^{\prime}=B_{\infty}^{\prime}
=\Phi_{\infty}^{\prime}=C_{\infty}^{\prime}=0$, $A_{\infty}=1$. 
The values of $\Phi_{\infty}$ and $B_{\infty}$ are not determined by the
field equations, and we next show how they may each be
rescaled to take on any value. 

The field equations (\ref{dAdx}) to (\ref{ddCdxx})
are invariant under the global rescaling
\be\label{rescale2}
   B\ra\gamma B,\;\;\;\varpi\ra\sqrt{\gamma}\varpi,\;\;\;
   C\ra\sqrt{\gamma}C,\;\;\;\Omega\ra\sqrt{\gamma}\Omega,
\ee
where $\gamma$ is some constant. Equation (\ref{rescale2}) simply
rescales the magnitude of the time-like Killing vector 
field, and leaves all solutions physically unchanged. With the gauge
freedom allowed by Eq.\  (\ref{rescale2}), the field equations 
become eigenvalue equations for $B_{0}$ with $\Omega$ arbitrary and
their solution in general leads to a value of $B_{\infty}\neq 1$. We
use Eq.\   (\ref{rescale2}) to set $B_{\infty}=1$ for each solution.

The equations of motion are also invariant under the global rescaling
\be\label{rescale1}
   \Phi\ra k^{2}\Phi,\;\;\; \sigma\ra k\sigma,\;\;\;
   C\ra k C,\;\;\; q\ra\frac{q}{ k},\;\;\;\Lambda\ra\frac{\Lambda}{k^{2}},
\ee
where $ k$ is some constant, provided $\omega(\Phi)$ is held
invariant. For a general ST theory, this will require a change in the
functional form of $\omega$. 

Equation (\ref{rescale1}) leaves the ratio
$S/\Phi_{\infty}$ and the ADM, Keplerian, Tensor and Newtonian masses
invariant and rescales the total charge and particle number as
\be\label{QNrescale1}
   Q\ra k Q,\;\;\;\; N\ra k^{2}N.
\ee
Physically, the rescaling exchanges gravitating
matter (measured by the number of bosons) for gravitational field
energy (which depends qualitatively upon the gravitational coupling
strength $G^{*}=G/\Phi$) in such a way as to leave the total
mass of the boson star invariant.

Consider now the set of solutions
\be\label{solnset1}
   {\cal S}(\sigma_{0};\Phi_{\infty},q,\Lambda)=
   \{M(\sigma_{0};\Phi_{\infty},q,\Lambda),\;
   N(\sigma_{0};\Phi_{\infty},q,\Lambda),\;
   Q(\sigma_{0};\Phi_{\infty},q,\Lambda)\}
\ee
parameterised by $\sigma_{0}$ for fixed $\Phi_{\infty}$, $q$ and 
$\Lambda$, where $M$ represents all of the mass values discussed in 
Section \ref{BV}. Under the rescaling $\Phi_{\infty}\ra
k^{2}\Phi_{\infty}$ the mass curves are re-parameterised as
\be
   M(\sigma_{0};\Phi_{\infty},q,\Lambda)\ra
   M(k\sigma_{0};k^{2}\Phi_{\infty},k^{-1}q,k^{-2}\Lambda)
\ee
while the particle number and charge curves are both re-parameterised
and rescaled as
\be
   N(\sigma_{0};\Phi_{\infty},q,\Lambda)\ra
   k^{2}N(k\sigma_{0};k^{2}\Phi_{\infty},k^{-1}q,k^{-2}\Lambda)
\ee
\be
   Q(\sigma_{0};\Phi_{\infty},q,\Lambda)\ra
   kQ(k\sigma_{0};k^{2}\Phi_{\infty},k^{-1}q,k^{-2}\Lambda).
\ee
Hence, by applying the rescaling (\ref{rescale1}), ${\cal S}$ may be
used to generate a new physically distinct set of solutions
\begin{eqnarray}\label{solnset2}
   {\widetilde {\cal S}}(\sigma_{0}; k^{2}\Phi_{\infty},q,\Lambda)
   & = & \{{\widetilde M}(\sigma_{0};k^{2}\Phi_{\infty},q,\Lambda),\;
   {\widetilde N}(\sigma_{0};k^{2}\Phi_{\infty},q,\Lambda),\;
   {\widetilde Q}(\sigma_{0};k^{2}\Phi_{\infty},q,\Lambda)\}
    \nonumber \\
   & = &\{M(k^{-1}\sigma_{0};\Phi_{\infty},kq,k^{2}\Lambda),
   \; k^{2}N(k^{-1}\sigma_{0};\Phi_{\infty},kq,k^{2}\Lambda),
   \; kQ(k^{-1}\sigma_{0};\Phi_{\infty},kq,k^{2}\Lambda)\}.
\end{eqnarray}
Thus, Eq.\   (\ref{rescale1}) may be used to
compare boson stars in cosmological settings with different values of
$\Phi_{\infty}$. 

The existence of this scaling relation means that we need to vary at
most two of the parameters $\Phi_{\infty}$, $\Lambda$ and $q$ in order
to investigate completely the properties of a solution in any given
theory. For example, by fixing $\Phi_{\infty}$ to some constant, the
complete set of solutions may be generated by the subset for which only
$\Lambda$ and $q$ are varied. In addition, if either of these 
two quantities is zero, we need only vary the other to build up a 
complete set of solutions. This will be of use when we discuss our
numerical results in Section \ref{NS}. There we shall be primarily
concerned with stars for which $\Lambda=0$ and we shall show complete
sets of solutions for several theories by only considering different
values of $q$.

Finally, we comment on another possible choice of mass and how it
is affected by Eq.\   (\ref{rescale1}).
Torres, Schunck \& Liddle \cite{TSL} {\it define} the Schwarzschild mass by
the relation
\be
   A=\left(1-\frac{2M^{*}}{x\Phi_{\infty}}\right)^{-1}.
\ee
Note that $A(x)$ is the same metric potential
both here and in Ref. \cite{TSL}.
Comparing this expression with Eq.\  (\ref{defMS}) we have $M^{*}=
\Phi_{\infty}M$, which implies that $M^{*}$ rescales by a
factor of $k^{2}$ under Eq.\  (\ref{rescale1}) with $q=0$. This
will be important when we analyse the behaviour of $M^{*}$ in a star
undergoing gravitational evolution (see Section \ref{GE}). They then 
define the binding energy by using
$M^{*}_{\infty}=\lim_{x\to\infty}M^{*}$ together with the rest 
mass (as we have defined it) to give
\be
   {\cal E}^{*}=M^{*}_{\infty}-N,
\ee
which also rescales by a factor of $ k^2$ under
Eq.\  (\ref{rescale1}). Their choice is such that
the binding energy of any solution has 
the same sign when compared with ours
and the fractional binding energy (defined as ${\cal B}={\cal
E}^{*}/N$ in \cite{TSL}) is identical for both papers. Thus, if
\be
   M^{*}_{\infty}-N<0
\ee
then
\be
   M_{\infty}-M_{N}\;\;=\;\;\frac{M^{*}_{\infty}}{\Phi_{\infty}}-M_{N}
   \;\;=\;\; \frac{M^{*}_{\infty}}{\Phi_{\infty}}-
   \frac{N}{\Phi_{\infty}}<0
\ee
and
\be
   \frac{M^{*}_{\infty}-N}{N}\;\;\equiv\;\; \frac{M_{\infty}-M_{N}}{M_{N}}.
\ee
Here we have used the ADM mass (as we have defined it) instead of the
Tensor mass since Torres et al.\  use the former in \cite{TSL}. 
As they point out, for
the large values of $\omega$ they consider, the numerical difference 
between $M_{ADM}$ and $M_{T}$ is negligible.

Since $M^{*}$ and the masses we define in this paper are measured in the same
units, it is apparent that they measure physically different quantities.
As mentioned in the text after Eq.\  (\ref{defMS}), the masses we define
implicitly include a factor of $1/\Phi_{\infty}$. This means that $M_{T}$
(or $M_{ADM}$ when $\omega$ is very large) measures the total
gravitational energy and $M_{K}$ measures the active gravitational
mass as seen by an orbiting test particle with negligible
self-gravity. Our mass definitions correspond to
the product $G{\cal M}$ of Newtonian theory, where ${\cal M}$
is the Newtonian rest mass. The Newtonian analogue of Eq.\
(\ref{rescale2}) is ${\cal M}\ra k^{2}{\cal M}$, $G\ra G/k^{2}$ which
leaves $G{\cal M}$ invariant: the rescaling swaps energy between the
matter and the gravitational field.  
In the weak field limit and when $\omega$ is large, 
$M^{*}_{\infty}$ describes the rest mass
of the star in which the gravitational coupling strength
$1/\Phi_{\infty}$ has been factored out (see the discussion after Eq.\
(\ref{wfcoupling}) in the following Section). Thus $M^{*}_{\infty}$
roughly corresponds to ${\cal M}$. However, for strong field solutions
one cannot easily interpret $M^{*}_{\infty}$, since one cannot
decouple the gravitational field strength from the total
energy. 

\section{Weak Field Limit and the Limiting Charge}\label{WF}

In this Section we discuss the weak field limit of the solutions which
we define as the limit in which $\sigma_{0}$ is small but non-zero.
The analysis that follows is in part a generalisation of the approach
followed by Kaup \cite{Kaup2} when dealing with uncharged boson stars in
GR and involves expanding the solutions in power series about flat
spacetime.  We start by defining a function $\Pi(x)$ which 
has the boundary value
$\Pi_{0}=1$ and the functional form given implicitly by
\be\label{Pidef}
   \sigma=\ve\Pi.
\ee
The constant parameter $\ve:=\sigma_{0}/\Pi_{0}$ measures the degree 
to which the
solution differs from flat spacetime and we shall use it as the
expansion parameter in the analysis. We also define a rescaled radial
coordinate by $a:=\sqrt{\ve}x$. Hence, $\Pi$ is now considered to be a
function of $a$. Note that, by definition, $\Pi(a)$ has the same
asymptotic form as $\sigma$ and decreases exponentially as $a\ra\infty$.

The field variables and energy parameter we expand in powers of $\ve$
about flat spacetime. We implicitly define new field 
variables $\alpha$, $\beta$, $\Sigma$ and $\Gamma$ by the first 
order expansions
\be\label{wfA}
   A=1+\ve\alpha(a)+{\cal O}(\ve^{2}),
\ee
\be\label{wfB}
   B=1+\ve\beta(a)+{\cal O}(\ve^{2}),
\ee
\be\label{wfS}
   \Phi=\Phi_{\infty}+\ve\Sigma(a)+{\cal O}(\ve^{2}),
\ee
\be\label{wfC}
   C=\frac{\ve}{2}\Gamma(a)+{\cal O}(\ve^{2})
\ee
and
\be\label{wfOM}
   \Omega=1-\frac{\ve}{2}E+\frac{\ve}{2}q\Gamma(a)+{\cal O}(\ve^{2}),
\ee
where $\Phi_{\infty}$ is the value of the scalar field at space-like
infinity and $E$ is a constant for each solution.
We assume that the scalar field coupling parameter has the form
\be\label{wfw}
   \omega=\wi+\ve W(a)+{\cal O}(\ve^{2}),\;\;\;
   \frac{d\omega}{d\Phi}=\Delta(a)+{\cal O}(\ve),
\ee
where $\wi$ is a constant,
$\Delta=\lim_{\ve\to 0}\left(\frac{d\omega}{d\Phi}\right)$ and
we have written the second term in the expansion for $\omega$
explicitly in terms of $a$. These assumptions restrict the applicability
of the following analysis to a subset of all possible scalar-tensor
theories. However, Eqs.\   (\ref{wfw}) are
satisfied for both BD theory and the power law theory. In the
latter case we have 
\be
   \omega=\frac{2n}{3\Phi_{\infty}^{n}}\left(\Phi_{\infty}+\ve\Sigma
   +{\cal O}(\ve^{2})\right)^{n}-\frac{3}{2}.
\ee
Expanding this expression binomially, we see that the power law theory has
$W=2n^2 \Sigma/(3\Phi_{\infty})$ and $\Delta=2n^2/(3\Phi_{\infty})$.

In terms of the new variables the field equations
(\ref{dAdx}) to (\ref{ddCdxx}) become
\be\label{wfdadr}
   (a\alpha)^{\prime}=\frac{4\Pi^{2}a^{2}(\wi+1)}
   {\Phi_{\infty}(2\wi+3)}+{\cal O}(\ve),
\ee
\be\label{wfdbdr}
   \beta^{\prime}=\frac{\alpha}{a}-\frac{2}{\Phi_{\infty}}
   \frac{d\Sigma}{da}+{\cal O}(\ve),
\ee   
\be\label{wfddbdrr}
   \left(a^{2}\beta^{\prime}\right)^{\prime}
   =\frac{4\Pi^{2}a^{2}(\wi+2)}
   {\Phi_{\infty}(2\wi+3)}+{\cal O}(\ve),
\ee
\be\label{wfddpdrr}
   \left(a^{2}\Pi^{\prime}\right)^{\prime}
   =\Pi(E+\beta-q\Gamma)+{\cal O}(\ve),
\ee
\be\label{wfddsdrr}
   \left(a^{2}\Sigma^{\prime}\right)^{\prime}
   =\frac{-2\Pi^{2}a^{2}}{2\wi+3}+{\cal O}(\ve)
\ee
and
\be\label{wfddcdrr}
   \left(a^{2}\Gamma^{\prime}\right)^{\prime}
   =2q\Pi^{2}a^{2}+{\cal O}(\ve),
\ee
where in this Section a prime denotes $d/da$. 
We have derived Eq.\   (\ref{wfddbdrr}) by differentiating Eq.\ 
(\ref{wfdbdr}) and substituting in Eqs.\   (\ref{wfdadr}) and (\ref{wfddsdrr}).
Equations (\ref{wfA}) to (\ref{wfC}) taken together with the asymptotic
conditions (\ref{asymflat}) and (\ref{asymC}) imply that we 
must impose the boundary conditions
\be\label{wfbdry}
   \alpha_{\infty}=\beta_{\infty}=\Gamma_{\infty}=\Sigma_{\infty}=0.
\ee
Equations (\ref{wfddbdrr}), (\ref{wfddsdrr}) and (\ref{wfddcdrr})
imply that both $\beta$
and $\Gamma$ are strictly increasing functions of $a$ while
$\Sigma$ is strictly decreasing. Combining these properties with the
boundary conditions above we have  $\beta\leq 0$, $\Gamma\leq 0$ 
and $\Sigma\geq 0$, where the equalities are only true in the 
limit $a\ra\infty$. From Eq.\   (\ref{wfOM}), the requirement that
$\Omega_{\infty}=\varpi/m<1$ and the vanishing of $\Gamma_{\infty}$ 
imply that $E\geq 0$, where the equality is only true when $\Pi=0$.

Combining Eqs.\   (\ref{Qlimit}) and (\ref{wfddcdrr}) we have
\be\label{wfQlimit}
   Q=\half\sqrt{\ve}\lim_{a\to\infty}\left(a^{2}\frac{d\Gamma}{da}\right)
   =\sqrt{\ve}\integ q\Pi^{2}a^{2}\;da
   +{\cal O}\left(\ve^{\thrhalf}\right)
   :=\sqrt{\ve}q{\cal N}+{\cal O}\left(\ve^{\thrhalf}\right),
\ee
where we have defined
\be\label{wfnumber}
   {\cal N}:=\integ a^{2}\Pi^{2}\;da.
\ee
The quantity $\sqrt{\ve}{\cal N}$ is the rest mass (or particle
number) to lowest order in $\ve$. From the definition (\ref{defMN}) the
Newtonian mass in the weak field limit is given by
\be\label{wfNewtonianmass}
   M_{N}=\frac{\sqrt{\ve}{\cal N}}{\Phi_{\infty}}
   +{\cal O}\left(\ve^{\thrhalf}\right).
\ee
\mbox{}From Eqs.\   (\ref{Scharge}) and (\ref{wfddsdrr}), the scalar
charge is given by
\be\label{wfSlimit}
   S=\sqrt{\ve}\lim_{a\to\infty}\left(a^{2}\frac{d\Sigma}{da}\right)
   =-\sqrt{\ve}\integ \frac{2\Pi^{2}a^{2}}{2\wi+3}\;da
   +{\cal O}\left(\ve^{\thrhalf}\right)
   =-\frac{2\sqrt{\ve}{\cal N}}{2\wi+3}+{\cal O}\left(\ve^{\thrhalf}\right),
\ee
where we have used Eq.\   (\ref{wfnumber}) to obtain the final equality.
Integrating Eq.\   (\ref{wfdadr}) and using the definition (\ref{defMS}),
we obtain the weak field ADM mass
\be\label{wfMADM1}
   M_{ADM}=\frac{2\sqrt{\ve}(\wi+1){\cal N}}{\Phi_{\infty}(2\wi+3)}
   +{\cal O}\left(\ve^{\thrhalf}\right),
\ee
where again we have used Eq.\   (\ref{wfnumber}). Combining Eqs.\  
(\ref{wfSlimit}) and (\ref{wfMADM1}) with the definitions
(\ref{defMT}) and (\ref{defMK}) gives the weak field tensor mass
\be\label{wfMT1}
   M_{T}=\frac{\sqrt{\ve}{\cal N}}{\Phi_{\infty}}+{\cal O}(\ve)
\ee
and the weak field Keplerian mass
\be\label{wfMK1}
   M_{K}=\frac{2\sqrt{\ve}(\wi+2){\cal N}}{\Phi_{\infty}(2\wi+3)}
   +{\cal O}\left(\ve^{\thrhalf}\right).
\ee
Equations (\ref{wfMT1}) and (\ref{wfMK1}) show that the Keplerian and
Tensor masses may be written as the products
$M_{K}=\sqrt{\ve}G_{K}{\cal N}$ and $M_{T}=\sqrt{\ve}G_{T}{\cal N}$,
where the coupling strengths $G_{K}$ and $ G_{T}$ are given by
\be\label{wfcoupling}
   G_{K}=\frac{2(\wi+2)}{\Phi_{\infty}(2\wi+3)},\;\;\;\;
   G_{T}=\frac{1}{\Phi_{\infty}}.
\ee
Hence, the weak field solutions are Newtonian in the sense that the
Keplerian and Tensor masses, which are the active masses to which test
particles respond, may be decomposed as  products of the star's rest mass
and a coupling strength. In addition we have $M_{T}=M_{N}$ to lowest
order in $\ve$, which motivates the labelling of the quantity defined
in  Eqn.\  (\ref{defMN}) as the Newtonian mass.

To calculate the fractional binding energy, we need to 
write $M_{T}$ and $M_{N}$
up to order $\ve^{\thrhalf}$. Rewriting the integrals (\ref{intS}) and
(\ref{intM}) in terms of the weak field variables, one can show that
the ADM mass and scalar charge are given by
\begin{displaymath}
   M_{ADM}=\frac{2\sqrt{\ve}(\wi+1){\cal N}}
   {\Phi_{\infty}(2\wi+3)}\;da+\ve^{\thrhalf}\integ\;
   \frac{a^{2}}{\Phi_{\infty}}\left[\frac{\Pi^{\prime 2}(2\wi+1)}{2(2\wi+3)}
   +\frac{\Gamma^{\prime 2}}{8}-\frac{\Sigma^{\prime}\beta^{\prime}}{4}
   +\frac{\Sigma^{\prime 2}\wi}{4\Phi_{\infty}}\right.
\end{displaymath}
\be\label{wfMADM}
   \left.-\frac{\Delta\Sigma^{\prime 2}}{2(2\wi+3)}
   +\frac{\Pi^{2}}{(2\wi+3)}\left(\frac{2W}{2\wi+3}+
   (2\wi+5)(q\Gamma-E-\beta)-\frac{2\Sigma}{\Phi_{\infty}}
   \left(\wi+1\right)\right)
   \right]\;da
   +{\cal O}\left(\ve^{\fhalf}\right)
\ee
and
\begin{eqnarray}\label{wfSINF}
   S=\frac{-2\sqrt{\ve}{\cal N}}{2\wi+3}
   +\ve^{\thrhalf}\integ\;a^{2}\left[-\frac{2\Pi^{\prime 2}}{2\wi+3}
   -\frac{\Delta\Sigma^{\prime 2}}{2\wi+3}
   +\frac{\Sigma^{\prime}}{2}\left(\alpha^{\prime}-\beta^{\prime}\right)
   \right. \nonumber \\ \left.
   +\frac{2\Pi^{2}}{(2\wi+3)}\left(
   q\Gamma-E-\beta-\alpha+\frac{2W}{(2\wi+3)}\right)
   \right]\;da
   +{\cal O}\left(\ve^{\fhalf}\right).
\end{eqnarray}
Similarly, rewriting Eq.\   (\ref{intQ}) and using the definition 
(\ref{defN}) we have
\begin{eqnarray}\label{wfMN}
   M_{N}=\frac{\sqrt{\ve}{\cal N}}{\Phi_{\infty}}
   +\ve^{\thrhalf}\integ\frac{a^{2}\Pi^{2}}{2\Phi_{\infty}}\left(
   q\Gamma-\beta-E+\alpha\right)\;da
   +{\cal O}\left(\ve^{\fhalf}\right).
\end{eqnarray}
Combining Eqs.\   (\ref{defMT}), (\ref{wfMADM}), (\ref{wfSINF}) and
(\ref{wfMN}) with the expression (\ref{benergy}) for the fractional binding
energy we have
\be\label{wfBE1}
   {\cal B}=\frac{\ve}{{\cal N}}\integ a^{2}
   \left[\frac{\Sigma^{\prime 2}\wi}{4\Phi_{\infty}}
   -\frac{\Sigma^{\prime}\alpha^{\prime}}{4}+\frac{\Pi^{\prime 2}}{2}
   +\frac{\Gamma^{\prime 2}}{8}
   +\frac{\Pi^{2}}{2}\left(\frac{-\alpha(2\wi+1)}{2\wi+3}
   -\frac{4\Sigma(\wi+1)}{\Phi_{\infty}(2\wi+3)}\right)
   \right]\;da
   +{\cal O}\left(\ve^{2}\right).
\ee
We need to simplify this expression considerably. To do this we 
integrate Eq.\   (\ref{wfBE1}) by parts in order to
write each term as some
numerical factor times the same positive definite integral. We carry
out each partial integration separately below, noting that all surface
terms vanish due to the asymptotic conditions (\ref{wfbdry}) and the
fact that all of the fields are finite at $a=0$.

Using the wave equation (\ref{wfddpdrr}) one can show that
\be\label{Pint1}
   \integ a^{2}\Pi^{\prime 2}\;da=-\integ a^{2}\Pi^{2}
   (E+\beta-q\Gamma)\;da.
\ee
Integrating the product $a\Pi^{\prime}(a^{2}\Pi^{\prime})^{\prime}$
by parts and using Eq.\   (\ref{Pint1}) one can also show that
\be\label{Pint2}
   \integ a^{2}\Pi^{\prime 2}\;da=\half\integ a^{3}\Pi^{2}
   (\beta^{\prime}-q\Gamma^{\prime})\;da.
\ee
Integrating $a^{2}\Sigma^{\prime}\alpha^{\prime}$ by parts and using Eqs.\  
(\ref{wfdbdr}) and (\ref{wfddsdrr}) we have
\be\label{ASint}
   \integ a^{2}\Sigma^{\prime}\alpha^{\prime}\;da=\integ \frac{2\Pi^{2}a^{3}}
   {2\wi+3}\left(\beta^{\prime}+\frac{2\Sigma^{\prime}}
   {\Phi_{\infty}}\right)\;da.
\ee
Integrating both $a^{2}\Sigma^{\prime 2}$ and the product $a\Sigma^{\prime}
(a^{2}\Sigma^{\prime})^{\prime}$ by parts, using the wave
equation (\ref{wfddsdrr}) and combining the results one can show that
\be\label{Sint1}
   \integ a^{2}\Sigma^{\prime 2}\;da
   =-\integ\frac{4 a^{3}\Pi^{2}
   \Sigma^{\prime}}{2\wi+3}\;da
   =\integ\frac{2a^{2}\Pi^{2}\Sigma}{2\wi+3}\;da.
\ee
Similarly, one can also show that
\be\label{Cint1}
   \integ a^{2}\Gamma^{\prime 2}\;da
   =\integ 4 a^{3}\Pi^{2}q\Gamma^{\prime}\;da
   =-\integ 2 a^{2}\Pi^{2}q\Gamma\;da.
\ee
Using Eq.\   (\ref{wfdadr}) we have
\be\label{Aint}
   \integ a^{2}\Pi^{2}\alpha\;da=\integ a^{3}\Pi^{2}\left(
   \beta^{\prime}+\frac{2\Sigma^{\prime}}{\Phi_{\infty}}\right)\;da.
\ee 
Substituting Eqs.\   (\ref{Pint2}) to (\ref{Aint}) into our expression
for the fractional binding energy (\ref{wfBE1}) gives
\be\label{wfBE2}
   {\cal B}=\frac{\ve}{{\cal N}}\integ a^{3}\Pi^{2}
   \left[\frac{q\Gamma^{\prime}}{4}
   +\frac{1}{2\wi+3}\left(\frac{\Sigma^{\prime}}
   {\Phi_{\infty}}(\wi+2)-\frac{\beta^{\prime}}{4}(2\wi+1)\right)\right]\;da
   +{\cal O}\left(\ve^{2}\right).
\ee
To simplify this expression further, we use Eq.\  (\ref{wfddcdrr}) to
obtain
\be\label{Cint3}
   \integ a^{3}\Pi^{2}\Gamma^{\prime}
   =\integ a\Pi^{2}\left(a^{2}\Gamma^{\prime}\right)\;da
   =\integ 2qXa\Pi^{2}\;da,
\ee 
where we have defined a new function
\be\label{Xdef}
   X(a):=\int_{0}^{a} {\tilde a}^{2}\Pi^{2}\;d{\tilde a}.
\ee
Similarly, from Eqs.\   (\ref{wfddbdrr}) and (\ref{wfddsdrr}) we have
\be\label{Bint1}
   \integ a^{3}\Pi^{2}\beta^{\prime}\;da=\integ\frac{4Xa\Pi^{2}(\wi+2)}
   {\Phi_{\infty}(2\wi+3)}\;da
\ee
and
\be\label{Sint3}
   \integ a^{3}\Pi^{2}\Sigma^{\prime}\;da=
   -\integ\frac{2Xa\Pi^{2}}{2\wi+3}\;da.
\ee
The quantity $\Pi^{2}$ is positive or zero over the interval of the
integration which implies that
$X$ is an increasing positive function of $a$ that vanishes at $a=0$. The
exponential decrease of $\Pi$ in the limit $a\ra\infty$ implies that $X$
has a finite limit $X_{\infty}$. Hence the integrals  (\ref{Cint3}), 
(\ref{Bint1}) and (\ref{Sint3}) are finite and substituting them
into Eq.\  (\ref{wfBE2}) gives
\be\label{wfBE3}
   {\cal B}=\frac{\ve}{{\cal N}}\left(\frac{q^{2}}{2}-\frac{\wi+2}
   {\Phi_{\infty}(2\wi+3)}\right)\integ aX\Pi^{2}\;da
   +{\cal O}\left(\ve^{2}\right).
\ee
This result is independent of $\Lambda$ and is true for any form of
the coupling parameter $\omega(\Phi)$ that satisfies Eq.\   (\ref{wfw}).

Equation (\ref{wfBE3}) implies that the fractional
binding energy is negative provided 
\be\label{qmax}
   q<q_{max}(\wi,\Phi_{\infty})=
   \sqrt{\frac{2(\wi+2)}{\Phi_{\infty}(2\wi+3)}}\;.
\ee
For the star to be stable ${\cal B}$ must be negative. Hence
Eq.\   (\ref{qmax}) places an upper limit on the charge to mass ratio of
the bosons that form a stable star in the weak field limit. For a star
with $q>q_{max}$ the Coulomb repulsion dominates over the
gravitational attraction and no stable solution exists. Note that, as
one would intuitively expect, $q_{max}\propto \sqrt{1/\Phi_{\infty}}$ 
for fixed $\omega$, so that the greater the
gravitational coupling strength, the greater the amount of charge that
can be bound into a stable object. In Section \ref{NS} we shall 
numerically examine how $q_{max}$ is modified for strong field
solutions. 

For finite $\omega$ we have $q_{max}>1$ for $\Phi_{\infty}=1$ which 
implies that in ST gravity it is possible to construct a stable 
star from bosons whose charge is greater than that allowed by GR.
In the GR limit, where $\omega\ra\infty$, Eq.\   (\ref{qmax}) 
gives $q_{max}=1$, which is the upper limit on $q$ given by Jetzer and van der 
Bij by just comparing the electrostatic and electromagnetic forces
\cite{JVB} (note that our unit of charge differs from theirs 
by a factor of $\sqrt{2}$). Hence, the above derivation serves 
also as a detailed proof of the maximum charge in General Relativity.

\section{Numerical Solutions}\label{NS}

In this Section we discuss strong field solutions of the field
equations (\ref{dAdx}) to (\ref{ddCdxx}) for both BD theory and 
the power law ST theory. In their full form,
the equations are complicated enough to
require numerical integration.
To generate each solution in the set ${\cal S}$, the numerical routine
that finds the eigenvalue $B_{0}$ must be nested inside a second
iterative routine that searches for the the value of $\Phi_{0}$ that
gives a value of $\Phi_{\infty}$ common to all solutions in ${\cal S}$.
When generating each solution, the numerical routine halts at the
value of $x$ at which $\sigma$ becomes too small for the code to
proceed further, typically at a value $\sigma\sim 10^{-10}$. Then the
Tensor mass is calculated using Eq (\ref{MTorbit}) with the limit
replaced by values of $x$ near to the termination point of the
routine and a power series approximation about $x=\infty$ is used 
to estimate $M_{T}$ for the solution. A similar power series method is
used to calculate $N$ and $S$, based on the behaviour of $\Phi(x)$
and the integral of $dN/dx$ out to the termination point.

For the sake of ease of comparison, we have chosen $\Phi_{\infty}=1$
for all of the solution sets we discuss here, and this choice implies
that $M_{N}$ and $N$ are numerically identical for each solution.
The sets of equilibrium solutions are shown in Figures 1 to 5.
In each of these Figures we show Tensor mass and Newtonian mass
curves as a function of $\sigma_{0}$ for several sets of solutions
within the same theory, choosing a different value of $q$ for each
pair of curves. The results of the numerical integrations are then used to
construct Figures 6 to 9. We shall discuss each figure separately
below.

Figure 1 shows the mass curves for several sets of solutions in
the $\omega=500$, $\Lambda=0$ BD theory, for 
which the weak field charge limit 
is $q_{max}=1.001$. As one would expect from such a
large value of $\omega$, the curves in Figure 1 are virtually
indistinguishable from the mass and particle number curves calculated
for GR boson stars in \cite{JVB}. As $q$ approaches $q_{max}$ from
below, the locations of the maxima in the $M_{T}$ and $M_{N}$ curves
shift to lower values of $\sigma_{0}$ and the binding energies ${\cal
E}$ decrease in magnitude, while the gradients $dM_{T}/d\sigma_{0}$
and $dM_{N}/d\sigma_{0}$ of the curves near $\sigma_{0}=0$
increase. In addition,
as $q$ increases, the mass curves shift to successively higher
values, since a star consisting of bosons with large $q$
must generate a strong gravitational field to overcome its own Coulomb
repulsion. This is achieved by an increase in $N$ (hence an increase in
$M_{N}$) which leads to an increase in $M_{T}$. Since
$\omega$ is large for these solutions, $\Phi$ 
remains approximately homogeneous throughout each
star in each solution set shown in the Figure. For example, in the
maximum mass $q=0.99$ solution, $\Phi_{0}$ and $\Phi_{\infty}$ differ
by less that $1\%$. Hence the solutions in the Figure show no strong
scalar-tensor gravitational effects.

In Figure 2 we show mass curves for several sets of solutions in the
$\omega=-1$, $\Lambda=0$ BD theory. As mentioned in Section \ref{INT}, this
choice of the coupling constant may be relevant to the study of boson
stars in the very early Universe, although in this case our value of
$\Phi_{\infty}=1$ is physically unrealistic: to approximate a boson
star embedded in the early (possibly string theory dominated) Universe, 
we should choose a value $\Phi_{\infty}<1$. However, from the
scaling relation (\ref{rescale1}), rescaling the asymptotic value of
$\Phi$ merely rescales the horizontal axis of Figure 2 and leaves the
form of the curves invariant (although for $\Phi_{\infty}\neq 1$,
$M_{N}$ and $N$ are no longer numerically equal and $q$ must be
rescaled). Hence the figure may be used as a template to generate
$\omega=-1$ boson stars solutions with any value of $\Phi_{\infty}$.

For the $\omega=-1$ BD theory, the weak field charge limit is
$q_{max}=\sqrt{2}$. As Figure 2 shows, this limit does not extend to
the strong field solutions (in contrast with the GR charged
boson stars for which the limit $q_{max}=1$ holds for all values of
$\sigma_{0}$). The mass curves of the solution set for which
$q_{max}>q>q_{c}$, where $q_{c}$ is some critical value of $q$,
diverge at finite values of $\sigma_{0}$, and the value of
$\sigma_{0}$ at which this happens decreases with increasing
$q$. The mass curves of the solution set satisfying $q<q_{c}$ do not
diverge, although for the values of $q$ presented here the value of
$\sigma_{0}$ at which the mass curves are maximal increases with
$q$. We have numerically found that $q_{c}\approx 0.90$ for the
$\omega=-1$ solutions. This value is approximate since the equations
are very hard to integrate when $q$ is close to $q_{c}$.

Physically, we may understand the reason for the divergent
behaviour of the masses as follows. For each solution, the ratio
$\Phi_{0}/\Phi_{\infty}>1$ and this quantity increases with
$\sigma_{0}$ since $\Phi$ is strongly coupled to the curvature.
Hence, within any particular solution set satisfying
$q>q_{c}$, the active
gravitational mass of each boson (measured
qualitatively by the product $m\Phi^{-1}$) decreases as $\sigma_{0}$ 
increases. At some finite $\sigma_{0}$, the number of bosons with charge
$q$ needed
to generate a gravitational field sufficient to overcome their own Coulomb
repulsion diverges. The results shown in Figure 2
imply that the limiting charge
$q_{max}$ is a decreasing function of $\sigma_{0}$ that has the weak
field value in the limit $\sigma_{0}\ra 0$ and asymptotes towards $q_{c}$ as
$\sigma_{0}\ra\infty$. Solutions with $q>q_{max}(\sigma_{0})$ do 
not exist, just as in the GR case.

Figure 3 shows sets of solutions for $\omega=1$
BD gravity with $\Lambda=0$. The weak field charge limit in this case
is $\sqrt{1.2}$. The coupling between $\Phi$ and the curvature
is weaker than it is in the $\omega=-1$ theory and the mass
curves show behavior intermediate between those in Figures 1 and
2. For values of $q$ up to $q\approx 0.9$, the location of the maxima
in the $M_{T}$ and $M_{N}$ curves decreases with increasing $q$. As
$q$ increases further, the location of the maxima starts to shift
towards higher values of $\sigma_{0}$ until the mass curves diverge at
$q=q_{c}\approx 0.976$. This value is smaller than the $\omega=-1$
critical charge and the numerical solutions suggest that $q_{c}$ is
also a decreasing function of $\omega$ for fixed $\sigma_{0}$ that
only approaches the GR charge limit as $\omega\ra\infty$.

Figure 4 shows mass curves for the power law ST theory, where
$2\omega(\Phi)+3=\frac{4}{3}n\Phi^{n}$. We have set $n=4$ in these
solutions; this parameter choice is in agreement with current 
observational constraints on this kind of 
ST theories \cite{Torres2}. These are
based on primordial nucleosynthesis calculations,
which for this theory are stronger
than the constraints imposed by solar system tests. (In fact, 
weak field test do not impose any bound on the exponent of the power 
law. This is a consequence of the form of the matter 
era cosmological solutions \cite{BARROW-M}, which are such that
$\omega\ra \infty$ and $\omega^{-3}d\omega/d\Phi \ra 0$ when
$t \ra \infty$ for all $n$.)
As with the BD
solutions, the mass curves diverge for values of $q$ greater 
than the critical charge, which for this theory is $q_{c}\approx
0.986$. The $n=4$ power law boson stars could conceivably be observed
today, and it is interesting that their mass curves are very different
{}from their GR counterparts.

We next outline how one converts from the dimensionless charge unit
we use here to the SI unit of charge. Rewriting the weak field
relationship (\ref{qmax}) as an equation that balances gravitational
and electrostatic forces on a boson in the asymptotic region of the
star, one finds that the charge ${\widetilde e}$ of each boson,
measured in coulombs, is given by
\be
   {\widetilde e}=qm\sqrt{\frac{4\pi\epsilon G}{\Phi_{\infty}}}
\ee
where $\epsilon$ is the permitivity of the boson star.
For a star of particle number $N$, the total charge ${\widetilde Q}$,
measured in coulombs, is
\be\label{qcoulombs}
   {\widetilde Q}=qm\sqrt{\frac{4\pi\epsilon G}{\Phi_{\infty}}}
   N\left(\frac{M_{pl}}{m}\right)^{2}.
\ee
For example, if we take $m=30 GeV\approx 5.3\times 10^{-26}$ kg, 
then ${\widetilde e}\approx (4.5\times 10^{-36})q$ C. For a boson star with
$N=q=1$, one has ${\widetilde Q}\approx 1$ C. This charge is
small. However, such a star will have a very small radius
(approximately $10^{-17}$ m) so that its charge density is very large.

In all of these solutions we have found that, for a given $N$ and $q$,
the magnitude of the scalar charge $S$ increases while the mass
$M_{T}$ decreases,
as the coupling strength $1/\omega$ is increased. This implies that for a
given number of bosons of charge $q$, the magnitude of the binding
energy ${\cal E}$ increases with the coupling strength. Thus the
bosons become more tightly bound even though the mass is smaller in
the strong coupling case. One can understand this apparent discrepancy
as follows. For all of these solutions, the scalar charge is negative
and, comparing Eqs. (\ref{defMT}) and (\ref{defMK}), this implies that
$M_{K}>M_{T}$, where the difference between these two quantities
increases with the coupling strength. It is even possible for $M_{K}$ to
exceed $M_{N}$ for particularly strong couplings 
(see \cite{AW} for an example). 
One can show that the fractional binding energies ${\cal B}$
are reasonably small for all of these solutions (even for the rather
extreme $\omega=-1$ solutions, for which ${\cal B}$ never exceeds $0.2$), so
that to a first order of approximation the bosons that make up the
atmosphere of a given star may be treated as a cloud of non self-gravitating
test particles. Here, the term atmosphere rather loosely describes the
asymptotic region of the star which encloses most of the mass,
electrical charge and scalar charge. However, as 
mentioned at the end of Section \ref{CM}, non self-gravitating 
test particles respond to the Keplerian mass. Hence, in
Newtonian terms, the gravitational force experienced by the bosons
in the atmosphere tends to increase as the coupling 
strength increases, even though the energy of the star decreases. For
a star with $q<q_{c}$, the bosons in the central region of the star
experience much the same conditions. The effects described here are
entirely separate to the weakening of the gravitational field strength
within the star caused by the increase of the ratio
$\Phi_{0}/\Phi_{\infty}$ with coupling strength. Although this reduces
the Keplerian mass, in a star with $q<q_{c}$ this effect is swamped by
the increase in the scalar charge. For stars with $q>q_{c}$, the
weakening of the gravitational field strength dominates over the
increase in scalar charge and the divergent behaviour discussed above
occurs. The fact that the scalar charge is large for strongly coupled
stars will be important when we discuss the scalarization phenomenon
below. 

Finally we consider the effect of introducing a quartic 
self-interaction term into the matter Lagrangian. For the GR solutions,
the inclusion of the self-interaction increases the mass of the stars
but does not alter the value of $q_{max}$ \cite{JVB}. We have found that for ST
theories, the inclusion of the $\Lambda$ term not only increases the
mass but also slightly decreases the charge limit of the strong field
solutions. To illustrate this, Figure 5 shows mass curves for the
$\omega=1$ BD theory in the limit $\Lambda\ra\infty$. The field
equations in this limit are obtained by making the change of variables
$\sigma\ra\sigma^{*}=\sigma\Lambda^{-1/2}$ and $x\ra
x^{*}=x\Lambda^{1/2}$, substituting these new variables into the
field equations (\ref{dAdx}) to (\ref{ddCdxx}), and taking the 
limit $\Lambda\ra\infty$ (see \cite{Colpi}, \cite{JVB} and \cite{Gund}
for details). The resulting field equations are identical
to Eqs.\  (\ref{dAdx}) to (\ref{ddCdxx}) except that the $\sigma^{\prime
2}$ terms vanish, $\sigma^{4}\Lambda$ is replaced by $\sigma^{*4}$,
$\sigma$ and $x$ are replaced by $\sigma^{*}$ and $x^{*}$
respectively, and Eq.\  (\ref{ddsdxx}) is replaced by the algebraic 
equation
\be
   \sigma^{*2}=\frac{\Omega^{2}}{B}-1.
\ee
The masses are now measured in units of
$\Lambda^{1/2}M_{pl}^{2}/m$, while $N$ is measured in units of
$\Lambda^{1/2}M_{pl}^{2}/m^{2}$. This implies that one must insert a
factor of $\Lambda^{1/2}$ in Eq.\  (\ref{qcoulombs}) to find the total
charge on these stars in SI units.

The curves in Figure 5 are qualitatively similar to
those in Figure 3 and the weak field value $q_{max}$ is the same as in
the $\Lambda=0$ case, since the limit imposed by Eq.\  (\ref{qmax})  
is independent
of $\Lambda$. However, the critical charge has decreased to a value
$q_{c}\approx 0.905$. This is to be expected, since a factor of
$\sigma^{4}\Lambda$ appears in the source term of the scalar wave
equation (\ref{ddsdxx}) which implies that a large value of $\Lambda$
leads to larger inhomogeneities in $\Phi$. Nevertheless, this effect is
still fairly minimal. Numerical calculations for other choices of
$\omega$ show similar behaviour.

Figures 6 and 7 are derived from the numerical output used to
generate Figure 4. Figure 6 is a bifurcation diagram for
the stars. The first five curves are for solutions with $q<q_{c}$ and
each shows a cusp at which the solutions become unstable. The
remaining three curves, with $q>q_{c}$, show no cusps and all are
stable. The other solution sets we have analysed have similar
diagrams and all of the solutions with $q>q_{c}$ are stable.
However, these objects have such large total charges and such small
binding energies that their formation may be halted by their own
Coulomb repulsion.

Figure 7 shows the behaviour of the coupling parameter $\alpha:=-S/M_{N}$
for the power law boson stars. Note that $\alpha$ here is different
{}from the weak field metric potential defined in Section \ref{WF}.
This coupling parameter measures the strength of the coupling
between the scalar field and the normal matter (in this case the
bosonic matter). In the GR limit, $\alpha\ra 0$. As the Figure shows,
this quantity is non-negligible even though the theory satisfies the
current observational constraints. 

The coupling parameter $\alpha$ was first introduced by Damour and
Esposito-Farese in the study of neutron star equilibrium
solutions \cite{SCALARIZATION}. 
They found that a phenomenon known as spontaneous
scalarization (SS) develops in the solutions: beyond a certain state
of compactness, a ST neutron star develops a non-trivial scalar field
configuration even when the coupling parameter $\omega$ of the ST
theory is chosen so that
$\omega\ra\infty$ far from the star. Their results are given in
Figure 2 of Ref.\  \cite{SCALARIZATION} and show that, when the
baryonic mass $M_{B}$ of the star is less that some critical value,
$\alpha$ is negligible, while as $M_{B}$ is increased beyond this value,
$\alpha$ rapidly grows to a maximum before decreasing again as $M_{B}$
is increased to its maximum value. $M_{B}$ is analogous to $M_{N}$
for bosons stars and Figure 7 shows that the power law boson stars
exhibit similar behaviour to neutron stars for large value of $M_{N}$
but differ considerably when $M_{N}$ is small. In this case, $\alpha$
increases towards some limiting value as $M_{N}$ decreases. This limit
may be calculated as follows. As $M_{N}\ra 0$, $\sigma_{0}\ra 0$ and
{}from the weak field equations (\ref{wfNewtonianmass}) and
(\ref{wfSlimit}) we have
\be\label{wfalpha}
   \lim_{\sigma_{0}\ra 0}\alpha=-\lim_{\sigma_{0}\ra 0}
   \left(\frac{S}{M_{N}}\right)=\frac{2\Phi_{\infty}}{2\omega_{\infty}+3}.
\ee
The difference between these two behaviours may be due to the
choice of the coupling parameter $\omega(\Phi)$. In Ref. 
\cite{SCALARIZATION}, this coupling has the form $2\omega+3\sim
1/(\log\Phi)$ which implies that, in the spatially asymptotic region,
$\omega\ra\infty$ for the choice $\Phi_{\infty}=1$. For the power law boson
stars we have $\omega_{\infty}=2/3\, n-3/2$ for 
$\Phi_{\infty}=1$, which is finite.

\section{Gravitational Evolution}\label{GE}

When we consider a boson star (or any other compact object)
in an evolving cosmological background,
we have to take into account the influence of the time varying
cosmological value of $\Phi$ on the structure of the star. To a
first approximation, and in want of a more rigorous analytical
analysis, we assume that $\Phi_{\infty}$ increases slowly with
cosmological time and that the boson star evolves quasi-statically so
that, at any point in its evolution, it is described by a static
equilibrium solution with the appropriate value of
$\Phi_{\infty}$. This approximation is a good one if the free fall
timescale of the star is much smaller than the characteristic
timescale over which $G^{\star}$ varies, and is often made when
considering other compact objects such as white dwarfs. Stars with
smaller values of $\Phi_{\infty}$ are assumed to exist at earlier
cosmological times,
since in a general ST cosmological model $\Phi$ tends to increase
with time. The star is modelled as a sequence of asymptotically flat
equilibrium solutions, where each star in the sequence has a value of
$\Phi_{\infty}$ greater than the preceding (earlier time) star. This
continues the approach adopted by Torres et al.\ 
\cite{TLS,TSL}, who showed how an increase in $\Phi_{\infty}$ affects
the physical characteristics of the star. They particularly discuss
the evolution of a star for which $\sigma_{0}$ remains constant,
although they also consider some other cases of more general evolution.

In this Section we discuss the possible gravitational evolution of 
both charged and uncharged boson stars of fixed particle number 
$N$. The assumption that $N$ is constant during
the evolution of an asymptotically flat configuration is justified
since in this case, Eq.\   (\ref{defN}) is valid and $N$ is a conserved
charge. However, in a cosmological setting (in which the spacetime is
not asymptotically flat), we cannot in general write a global
conservation law for any quantity, but instead must approximate a
quasi-local conservation equation. This problem is not unique
to ST boson stars in a cosmological setting: for any metric theory of
gravity the formulation of global conservation laws in a general
spacetime is very difficult. In this analysis we assume that the
asymptotic region of the boson star matches smoothly to a homogeneous
cosmological solution at some large value $x_{1}$ of the radial
coordinate where the boson field amplitude $\sigma$ is
negligible. We assume that the quasi-local value of the particle number
of the star
in a cosmological setting (evaluated as an integral out to the matching
surface $x=x_{1}$ over some space-like hypersurface) is approximately
equal to $N$, the particle number the star would have if isolated in an
asymptotically flat spacetime. Note that the
Jordan frame representation of ST gravity that we use in this paper
embodies the Einstein Equivalence Principle (defined in \cite{Will})
so that local charge conservation holds and $q$ is constant during the
star's evolution. Hence if $N$ is constant, this implies
that $Q$ remains constant. However, as we shall show below, $M_{T}$ is
not conserved during the evolution. 
Physically, the assumption that $N$
is constant is reasonable since classically, this quantity may be
literally interpreted as a count of the number of bosons in the star, a
procedure that is coordinate independent. The mass, on the other
hand, is treated as a quasi-local quantity when the star is in a
cosmological setting, and there are no conservation laws or physical
arguments to constrain its behaviour.

We first examine the evolution of an uncharged boson star of particle
number $N$ with $\Lambda=0$ in BD gravity. 
We make use of the notation of Section
\ref{BV} in which a set of solutions is denoted ${\cal
S}(\sigma_{0};\Phi_{\infty}, q, \Lambda)$, where $\Phi_{\infty}$, $q$
and $\Lambda$ are fixed and $\sigma_{0}$ is a parameter that labels
the different elements of ${\cal S}$. In the discussion that follows,
$\sigma_{0}$ is an independent variable while $\sigma_{1}$ and
$\sigma_{2}$ are particular values of this variable that label
particular solutions within the set.
  
Let the initial asymptotic value of the
scalar field be $\Phi_{1}=\Phi_{\infty}(t_{1})$ at some initial time
$t_{1}$ so that the star is a member of the set ${\cal
S}_{1}(\sigma_{0};\Phi_{1},q=0,\Lambda=0)$. We assume that the star is
initially stable, which implies that it lies to the left of the first
maximum in the $N(\sigma_{0})$ curve \cite{KMS}. Then the pair
$(N,\Phi_{1})$ uniquely determines the initial central
amplitude $\sigma_{1}$ of the boson field. The star will have a mass
$M_{1}=M_{T}(\sigma_{1},\Phi_{1},q=0,\Lambda=0)$. Here $\sigma_{1}$ is the
particular value of $\sigma_{0}$ that gives the element of ${\cal
S}_{1}$ with particle number $N$. At some later
time $t_{2}$, the asymptotic value of the scalar field has evolved to
$\Phi_{2}=\Phi_{\infty}(t_{2})=k^{2}\Phi_{1}$ where $k>1$. At this
time the star is a member of the set ${\cal
S}_{2}(\sigma_{0};\Phi_{2},q=0,\Lambda=0)$ with particle number
$N$. In general, this solution will have a central scalar field
amplitude $\sigma_{2}$ different from $\sigma_{1}$. From Eqs.\  
(\ref{solnset1}) and (\ref{solnset2}), the later time solution is 
generated by a rescaling of the member of ${\cal S}_{1}$ that 
has particle number $k^{-2}N$. Since $k>1$, and since 
for stable stars in ${\cal S}_{1}$ the particle
number curve $N(\sigma_{0})$ obeys $dN/d\sigma_{0}>0$,
this generating solution has
$\sigma_{2}<\sigma_{1}$. For stable solutions
$dM_{T}/d\sigma_{0}>0$, which implies that the mass
$M_{2}=M_{T}(\sigma_{2},\Phi_{1},q=0,\Lambda=0)$ of the generating
solution obeys the relation $M_{2}<M_{1}$. To find the
physical characteristics of the new solution at time $t_{2}$ we must
use Eq.\  (\ref{rescale1}) to rescale the parameters of the generating
solution using the rescaling factor $k$. However, $M_{T}$ is invariant
under this rescaling which implies that the later time boson 
star has a smaller mass than the earlier time solution.
Repeating this process for successively
higher values of $\Phi_{\infty}$ (or, equivalently, later cosmological
times), one can see that the physical properties of the star during
its evolution may be calculated from the single initial solution set
${\cal S}_{1}$: as the cosmological time increases, the generating
solution is located at an ever decreasing value of $\sigma_{0}$ and
its mass, as well as the mass of the physical solution it generates,
decreases. In addition, the scalar field becomes increasingly
homogeneous and its central value decreases, as viewed from the rest
frame of the star. Physically, the mass loss may be accounted for by the
generation of scalar gravitational radiation, which carries energy out
{}from the star, and the weakening of the gravitational coupling strength.
As pointed out in \cite{TSL}, within a single
solution set the radius of a stable boson star is an increasing
function of $1/\sigma_{0}$ (stars approach infinite radius in the
weak field limit $\sigma_{0}\ra 0$). The radius, like the mass, is 
invariant under Eq.\  (\ref{rescale1}), so that this quantity 
also increases with
cosmological time. Note that if the star starts out as being stable at
time $t_{1}$, then it will remain stable throughout its
evolution.

Figure 8 shows evolution curves for four stable, uncharged boson stars in
$\omega=-1$ BD gravity, where each star has a different particle
number $N$. Each star evolves from an initial equilibrium solution
with $\Phi_{\infty}$=1 to a configuration with $\Phi_{\infty}=6$. The
generating solutions all lie on the $q=0$ curves of Figure 2, and all
of the stars evolve towards lower values of $M_{T}$ and
$\sigma_{0}$ (in other words, the generating solutions approach the
weak field limit as the cosmological time increases). For the 
stable part of each set ${\cal S}$, the 
quantity $\sqrt{1-\Omega_{\infty}^{2}}$, which gives a
measure of the inverse radius of the star, tends towards $0$ as
$\sigma_{0}\ra 0$ (this is true for all of the solutions investigated
here). Hence the radius of each star increases as it evolves. If we
were to continue the evolution beyond
$\Phi_{\infty}=6$, we would find that in the limit
$\Phi_{\infty}\ra\infty$, the stars tend towards a state of zero mass,
zero boson field amplitude and infinite radius (i.\  e.\   they
disperse). Furthermore, since the ratio $\Phi_{0}/\Phi_{\infty}$
decreases towards unity as $\sigma_{0}$ decreases, the scalar field in
the interior of an evolving star becomes increasingly homogeneous.

We now compare these results with the evolution curves shown in Figure
1 of Ref.\  \cite{TSL}. There, it was found that the mass
$M^{*}_{\infty}$ of the
equilibrium solutions increased with time. This behaviour may be 
understood if we recall the discussion of Section \ref{BV}.
The definition of mass used in \cite{TSL}
differs from the one used here by a factor of $\Phi_{\infty}$. On
using the definition in the former paper, one finds that the mass
increases when we rescale from the generating solution (with
$\Phi_{\infty}=1$) to the physical solution 
(with $\Phi_{\infty}>1$).\footnote{We note here that the choice
$\Phi_{\infty}=1$ for the generating solution is for convenience only
and similar curves could be drawn for stars
evolving from the distant past (with $\Phi_{\infty}<1$) through to the
present cosmic time (with $\Phi_{\infty}=1$) by making a second
rescaling of the parameters of the physical solutions.} The mass we
use here is invariant under this rescaling. A
quantitative example of how these results differ for constant $N$
evolution is given in Table 1. 
The conclusions in \cite{TSL} are seen to be correct once one recognises that
their mass is physically different to the definition we adopt here. 

For the sake of completeness, we briefly outline how an uncharged star
of constant $\sigma_{0}$ evolves. This kind of evolution is discussed
in Ref.\  \cite{TSL} and we reproduce and expand on their results.
Since the rescaling factor $k$ increases with
cosmic time, the generating solutions must be those of ever decreasing
values of $\sigma_{0}$ that approach the weak field limit
as time increases. As is the case for the fixed $N$ evolution, the 
mass $M_{T}$ of the constant $\sigma_{0}$ stars decreases with
time. Here $N$ is no longer conserved and in the physical solution
will in general increase with time. This can be seen as follows. Under
a small change $\delta\sigma_{0}$ in the generating solution we have
$\sigma_{0}\ra{\widetilde \sigma}_{0}=\sigma_{0}+\delta\sigma_{0}$. 
Since $\sigma_{0}$ is constant in the physical solution, we have
$k=1-\delta\sigma_{0}/\sigma_{0}$ to first order in
$\delta\sigma_{0}/\sigma_{0}$ and, since $k>1$, this implies that
$\delta\sigma_{0}<0$ to lowest order. 
The particle number $N$ changes according to the relation $N\ra {\widetilde
N}=N+\delta\sigma_{0}(dN/d\sigma_{0}$), where ${\widetilde N}$
is the particle number of the generating solution at ${\widetilde 
\sigma}_{0}$ and we have written this expression to lowest order in 
$\delta\sigma_{0}\sigma_{0}$. Rescaling to find the corresponding 
particle numbers of
the physical solutions and taking their difference 
we have $k^{2}{\widetilde N}-N=\delta\sigma_{0}
(dN/d\sigma_{0}-2N/\sigma_{0})$, which is
positive wherever the generating solution satisfies the inequality
$(\sigma_{0}/N)\, dN/d\sigma_{0}<2$. One
can show numerically that this is true for all stable solutions with
$q<q_{c}$. For any particular set of solutions with $q>q_{c}$, the 
inequality is satisfied only up to some particular value of
$\sigma_{0}$ that varies with the parameters of the set.
Again, as discussed in Section \ref{BV}, the mass
$M^{*}_{\infty}$ used in Ref. \cite{TSL} is different from the one
used here, and it rescales by a factor of 
$k^{2}$ under Eq.\ (\ref{rescale1}). In a constant $\sigma_{0}$
evolution, one can show
that $M^{*}_{\infty}$ will be increasing with time
if the condition $(\sigma_{0}/M^{*}_{\infty})\,dM^{*}_{\infty}/
d\sigma_{0}<2$ holds for the generating solution. This is
true for all solutions with $q<q_{c}$ and for solutions with $q>q_{c}$
whose central boson field amplitude is below some particular
value. 

We next consider the evolution of a charged boson star with
$\Lambda=0$ in BD gravity. The situation
here is more complex since we must also deal with the charge
rescaling in Eq.\  (\ref{rescale1}). To start with, we consider a stable
boson star of particle number $N$ and mass $M_{1}$ made up of
bosons with charge $q$, where the asymptotic scalar field has the
value $\Phi_{1}$. The solution is a member of the set ${\cal
S}_{1}(\sigma_{0};\Phi_{1},q,\Lambda=0)$. The choice of $N$
and $q$ uniquely determines the central boson field amplitude
$\sigma_{1}$ of the solution. Again, keeping $N$ fixed we
evolve $\Phi_{\infty}$ to a new value $\Phi_{2}=\Phi_{\infty}(t_{2})=
k^{2}\Phi_{1}$ where
$k>1$. The star is now a member of the set ${\cal
S}_{2}(\sigma_{0};\Phi_{2},q,\Lambda=0)$ with particle number
$N$. However, in this case the new solution cannot be generated by
rescaling a solution in ${\cal S}_{1}$, since we must also take into 
account the rescaling of $q$. Instead the generator 
of the new solution is the member of the set ${\cal
S}_{3}(\sigma_{0};\Phi_{1},kq,\Lambda=0)$ with particle number
$k^{-2}N$. Repeating this process for successively higher values of
$\Phi_{\infty}$ we see that the generating solutions at each
cosmological time are those of successively higher values $q$. 

Figure 9 shows the
evolution of several charged boson stars in $\omega=-1$ BD theory. The
stars have charges $q=0.7$, $q=0.8$, $q=0.9$ and $q=1.0$, and each has
$N=1$ particles. Each of the solutions evolves from an initial value
of $\Phi_{\infty}=1$ to a final value of $\Phi_{\infty}=2.8$. 
The direction of increasing time in the Figure
is from top right to bottom left, and each star evolves to a
configuration of lower mass and lower central boson field amplitude. Some of
the $\Phi_{\infty}=1$ solutions used to generate the late time
solutions are shown as points in Figure 2, where later time
generating solutions are those of higher values of $q$. 
Their corresponding rescaled
(physical) solutions are shown as points in Figure 6. Unlike in the
uncharged case, charged boson stars cannot evolve indefinitely. At
some point in a star's evolution, its generating solution has a value
of $q$ that exceeds $q_{c}$ and the star will no longer be stable: as
the cosmological value of $G^{\star}$ becomes weaker (ie.\
$\Phi_{\infty}$ increases), $\Phi$ in the interior of the star 
becomes increasingly more homogeneous while its central value also
increases. Eventually, the gravitational field inside the star will
become too weak for the star to remain gravitationally bound and it
will be dispersed by its own Coulomb field after some finite
time. This phenomenon is unique to ST gravity, since it is only in
these theories
that the gravitational field strength in the star may weaken with time.

The evolution of a star in a more general ST theory is a little more
complex since $\omega$ will not be invariant under a rescaling of
$\Phi$ unless its functional form is changed. For example, given the
power law coupling considered above, one would also have to rescale
the numerical coefficient of $4/3$ by a factor of $k^{-2n}$. Hence, to
analyse the evolution of a ST star one would have to start with a
sequence of generating solutions, each with a different choice of
$\omega$, much as in the case of a charged BD stars in which we needed
a sequence of generating solutions, each of a different charge.

\section{Conclusions}\label{CONCLUSIONS}

In this work, we have studied the equilibrium configurations
of charged boson stars both in Brans-Dicke theory and scalar-tensor
gravity with a power law coupling, and we have re-examined uncharged
boson star solutions.  We have found that, theoretically,
these structures are stable and might exist in 
the real Universe, provided of
course that a ST theory correctly describes the gravitational
interaction.
However, it is likely that only rotating
stars will exist in an astrophysical setting. 
While we have made no attempt to construct solutions describing
rotating charged stars, the existence
of rotating stars in the uncharged case leads us to suppose that charged 
objects of this kind may also exist in GR and ST gravity. 
The use of charged objects to generate astrophysical effects was recently 
analysed for the case of black holes, and this constitutes an additional
motivation for the consideration of other charged stellar structures.
Additionally, the study of the range of
possible stellar structures in ST gravity
is far from being complete, and a careful understanding of the influence
of an evolving $G^\star$  in a cosmological background is still missing. 
Without stating again each of the results we have obtained in the previous
sections, we may say that the use of simple
relativistic objects, such as the boson stars explored here, provide a
useful setting in which to discuss gravitational issues and to compare ST
theories with GR. For example, we have found that the properties of both
static and evolving charged ST solutions in a cosmological
background differ considerably from those of GR solutions. By clarifying 
the mass definitions and the properties of the rescaling, 
we have shown for the first time that gravitational evolution may
drive a compact object towards a state of decreased energy and central
density. The degree to which our conclusions are true for more complex
and realistic astrophysical objects is unclear. However, it seems
likely that these objects 
will yield observable astrophysical signals, like those reported in 
Refs.\  \cite{BERRO,BENVE} for the case of white dwarfs.
Much more work has yet to be done,
particularly in the field of gravitational memory and gravitational
evolution.

\section*{Acknowledgments}

D.F.T. was supported by CONICET. He acknowledges The British Council
for awarding him with a Chevening Fellowship, which allowed this research
to start. We acknowledge profitable conversations with G. E. Romero (IAR),  
and F. E. Schunck (K\"oeln) during the course of this research, and earlier
conversations with A. R. Liddle (Imperial).

\newpage 

\begin{table}
\caption{
Here we show some data from an $\omega=-1$, $q=0$ BD boson star
evolution, in which the particle number is conserved. We have chosen
$N=0.6$. Columns 2, 3 and 4 show the parameters of the 
generating solutions: central boson field amplitude $\sigma_{0}$, 
particle number $N$ and mass (here we use the Jordan frame ADM mass
for ease of comparison). Columns 5 and 6 show $\sigma_{0}$
and the mass $M^{*}_{\infty}$ of the physical solution, using the 
mass definition adopted in the paper by Torres, Schunck and Liddle. 
Since this mass rescales by a factor $k^2$,
it is larger than in the generating solution and increases
with $\Phi_{\infty}$. In the present paper, the physical solution has equal
ADM mass to the generating solution, since $M_{ADM}$ is invariant under
the rescaling. Thus this mass decreases with cosmic time.}
\begin{tabular}{c|lll|ll}
$\Phi_{\infty}$ &  $\sigma_{0}$ &    $N$  &   $M_{ADM}$  &
$\sigma_{0}$ &  $M^*_{\infty}$\\
1.00000  &   0.242902 &  0.600000 &  0.520288 &    0.242902 &  0.520288\\
1.20000 &    0.159362 &  0.500000 &  0.446675 &    0.174572 &  0.536010\\
1.40000 &    0.117526 &  0.428571 &  0.393866 &    0.139059 &  0.551412\\
1.60000 & 9.14216E-02 &  0.375000 &  0.352852 &    0.115640 &  0.564563\\
1.80000 & 7.16179E-02 &  0.333333 &  0.315893 & 9.60855E-02 &  0.568607\\
2.00000 & 5.77638E-02 &  0.300000 &  0.285910 & 8.16904E-02 &  0.571820\\
\end{tabular}
\label{Table1}
\end{table}

\newpage

\begin{figure}
\centering 
\leavevmode
\epsfxsize=8cm 
\epsfbox{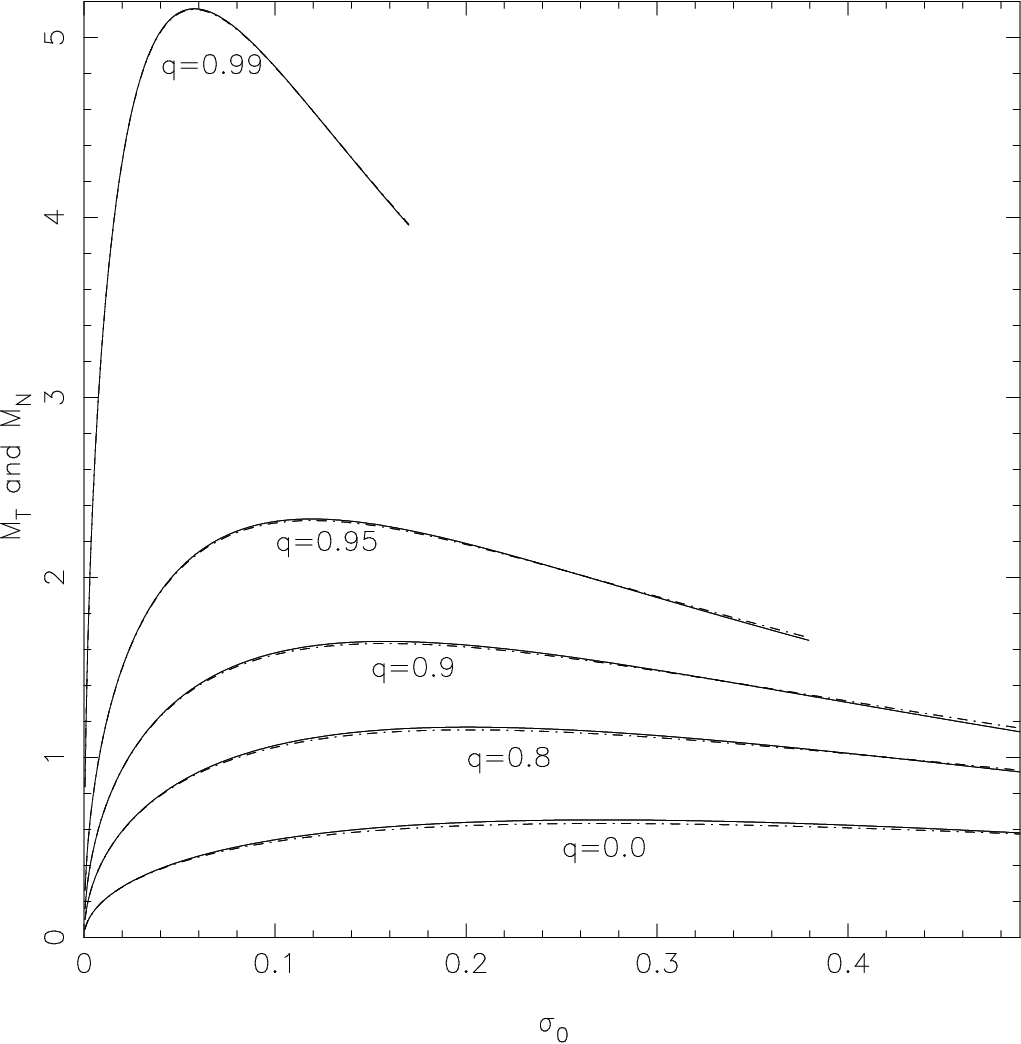}\\
\caption{Mass curves for the $\omega=500$ Brans Dicke
theory with $\Phi_{\infty}=1$ and $\Lambda=0$. The curves are 
parameterised by $\sigma_{0}$ and labelled by the boson charge-to-mass
ratio $q$. For each value of $q$ two curves are shown: the Tensor mass
$M_{T}$ (broken curves) and the Newtonian mass $M_{N}$ (solid
curves). Both masses are measured in units of $M_{pl}^{2}/m$. 
Our choice of $\Phi_{\infty}$ implies that $M_{N}$ is numerically equal
to the particle number $N$ and $M^*_{\infty}$ is numerically equal to
$M_{ADM}$.}
\end{figure} 

\begin{figure}
\centering 
\leavevmode\epsfxsize=8cm \epsfbox{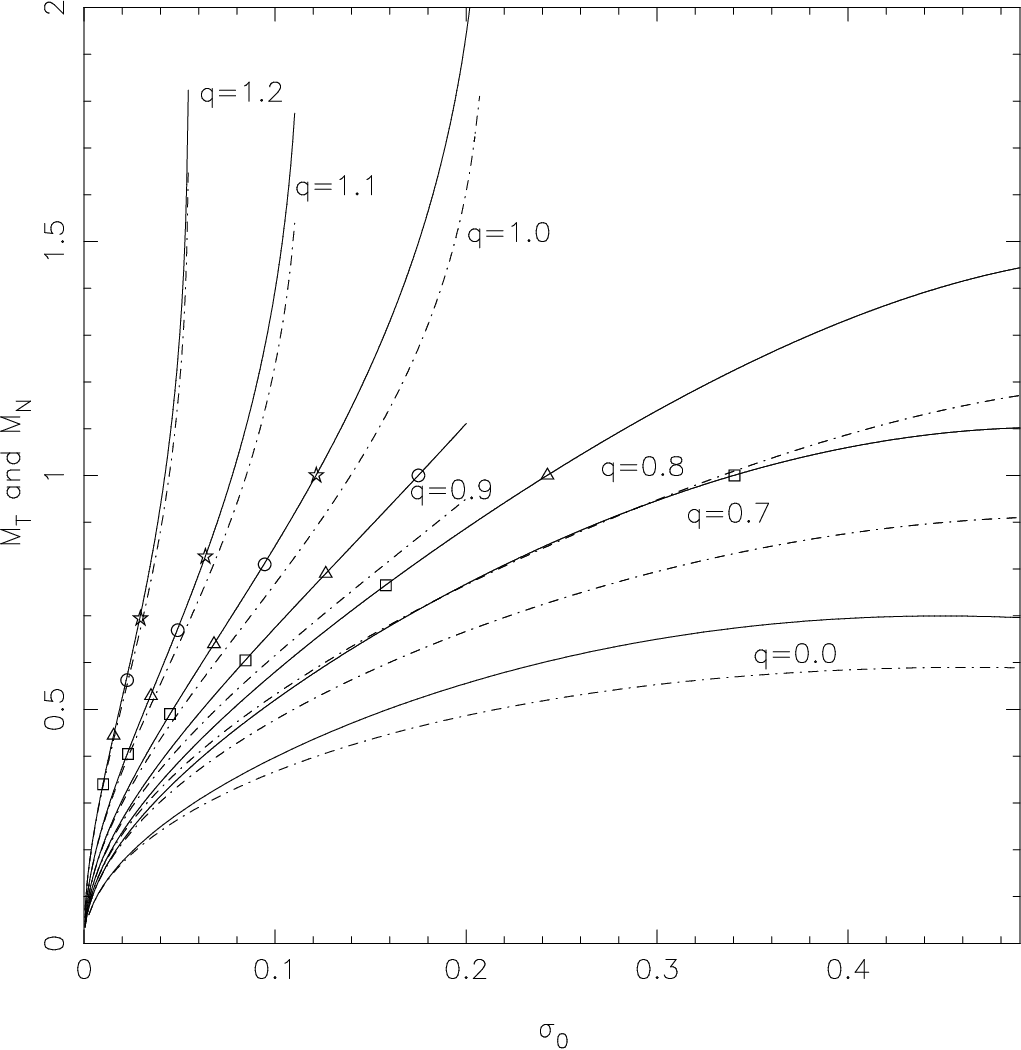}\\ 
\caption{Mass curves for the $\omega=-1$ Brans Dicke theory with
$\Phi_{\infty}=1$ and $\Lambda=0$. Curve labelling and parameterisation are
the same as in Figure 1. The weak field charge limit is
$q_{max}^2=2$  and the critical charge is $q_{c}=0.90$, above
which the mass curves diverge. The data points show solutions used to
generate the evolution sequences in Figure 9. Four sets of generating
solutions are shown: $q=1.0$ (marked by stars), $q=0.9$ (marked by
circles), $q=0.8$ (marked by triangles) and $q=0.7$ (marked by
squares). The initial solution for each evolution sequence is at
$M_N=1$ and the later time generating solutions are those of higher
values of $q$ and lower values of $M_N$, $M_T$ and $\sigma_0$. See the
text in Section VII for further explanation.}
\end{figure} 

\begin{figure}
\centering 
\leavevmode\epsfxsize=8cm \epsfbox{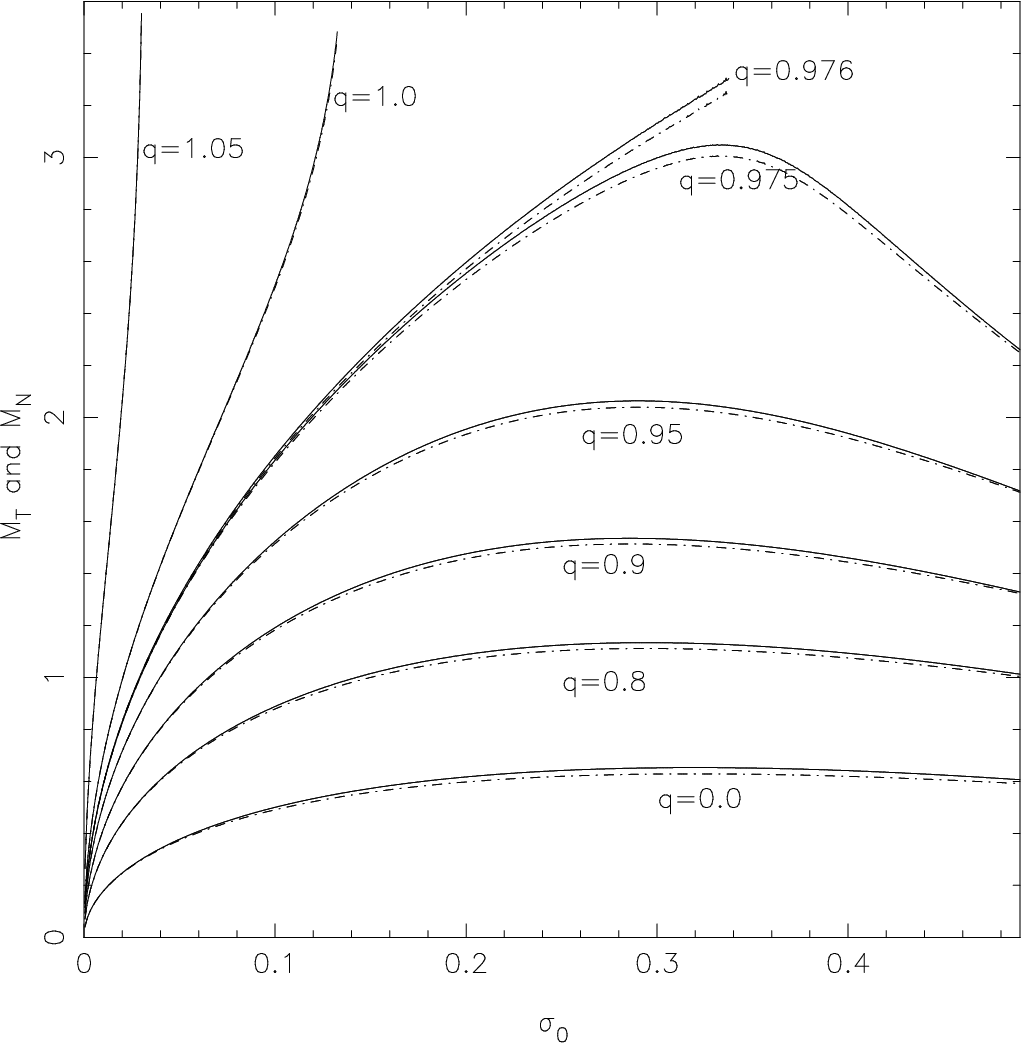}\\ 
\caption{Mass curves for the $\omega=1$ Brans Dicke theory with
$\Phi_{\infty}=1$ and $\Lambda=0$. Curve labelling and parameterisation are 
the same as in Figure 1. The weak field charge limit is
$q_{max}^2=1.2$  and the critical charge is $q_{c}=0.976$.}
\end{figure} 

\begin{figure}
\centering 
\leavevmode\epsfxsize=8cm \epsfbox{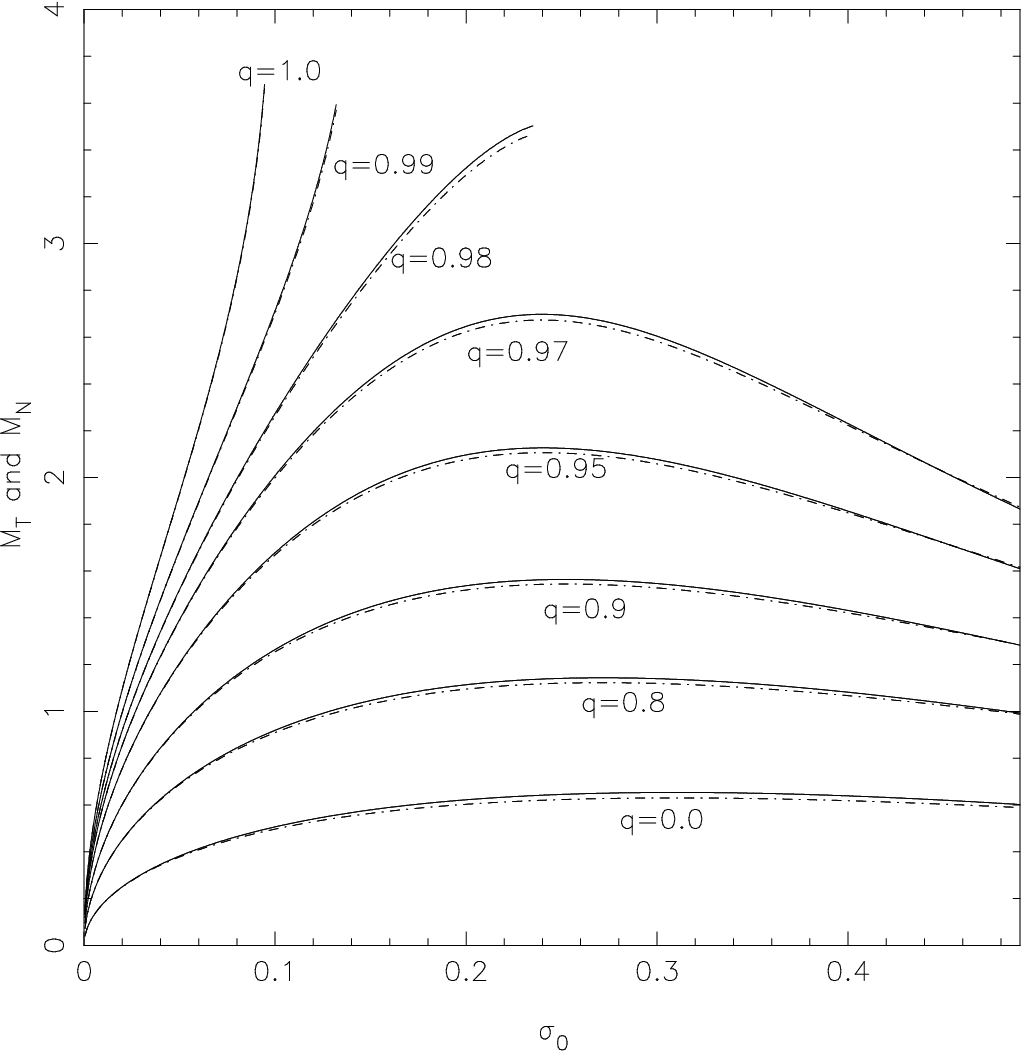}\\ 
\caption{Mass curves for the $2\omega+3=\frac{4}{3}n\Phi^{n}$ power
law ST theory with $n=4$, $\Phi_{\infty}=1$ and $\Lambda=0$. 
Curve labelling and parameterisation are the same as in Figure 1. 
The weak field charge limit is $q_{max}^2=19/16$  and the
critical charge is $q_{c}=0.986$.}
\end{figure} 

\begin{figure}
\centering 
\leavevmode\epsfxsize=8cm \epsfbox{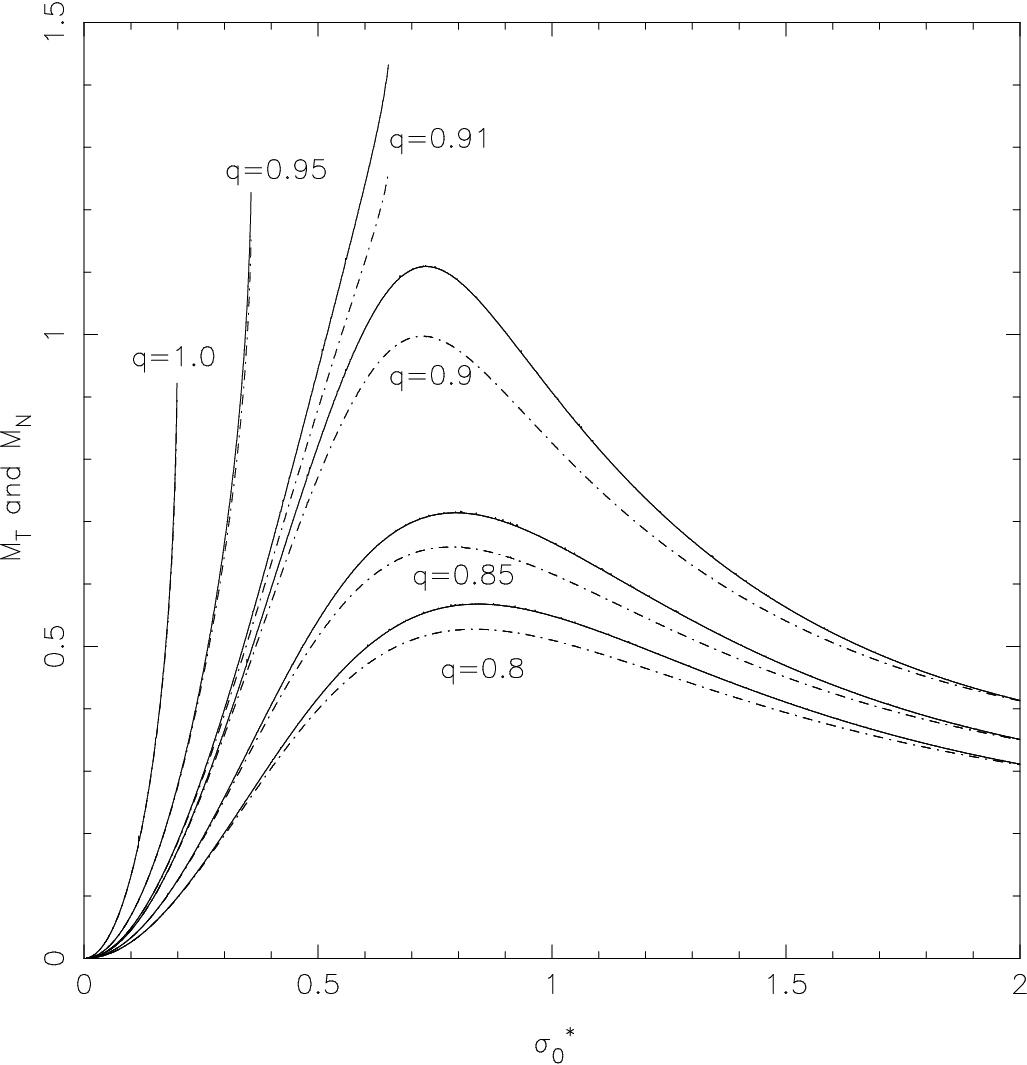}\\ 
\caption{Mass curves for the $\omega=1$ Brans Dicke theory in the
limit $\Lambda\ra\infty$ with $\Phi_{\infty}=1$. Curve labelling
is the same as in Figure 1. The curves are parameterised by the
rescaled central density $\sigma^{*}$ and the masses are measured in
units of $M_{pl}^{2}/(m\Lambda^{1/2})$. The weak field charge limit is
$q_{max}^2=1.2$ (the same as in Figure 3) while the critical
charge is $q_{c}=0.905$.}
\end{figure} 

\begin{figure}
\centering 
\leavevmode\epsfxsize=8cm \epsfbox{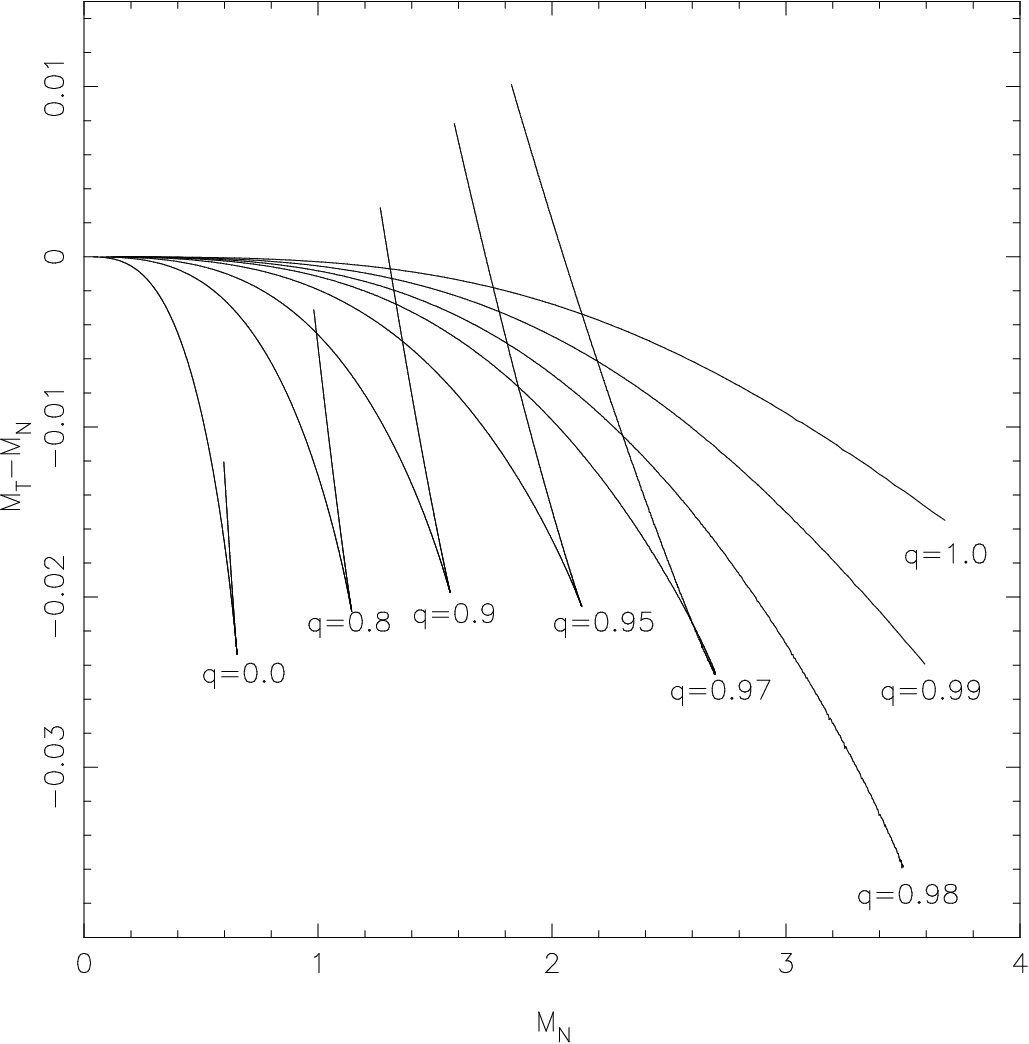}\\ 
\caption{Bifurcation diagram for charged boson stars in the $n=4$
power law theory. Using catastrophe theory one may show that 
the appearance of cusps
signals a change in the stability of the stellar object. Branches that
start at the coordinate origin are stable.}
\end{figure} 

\begin{figure}
\centering 
\leavevmode\epsfxsize=8cm \epsfbox{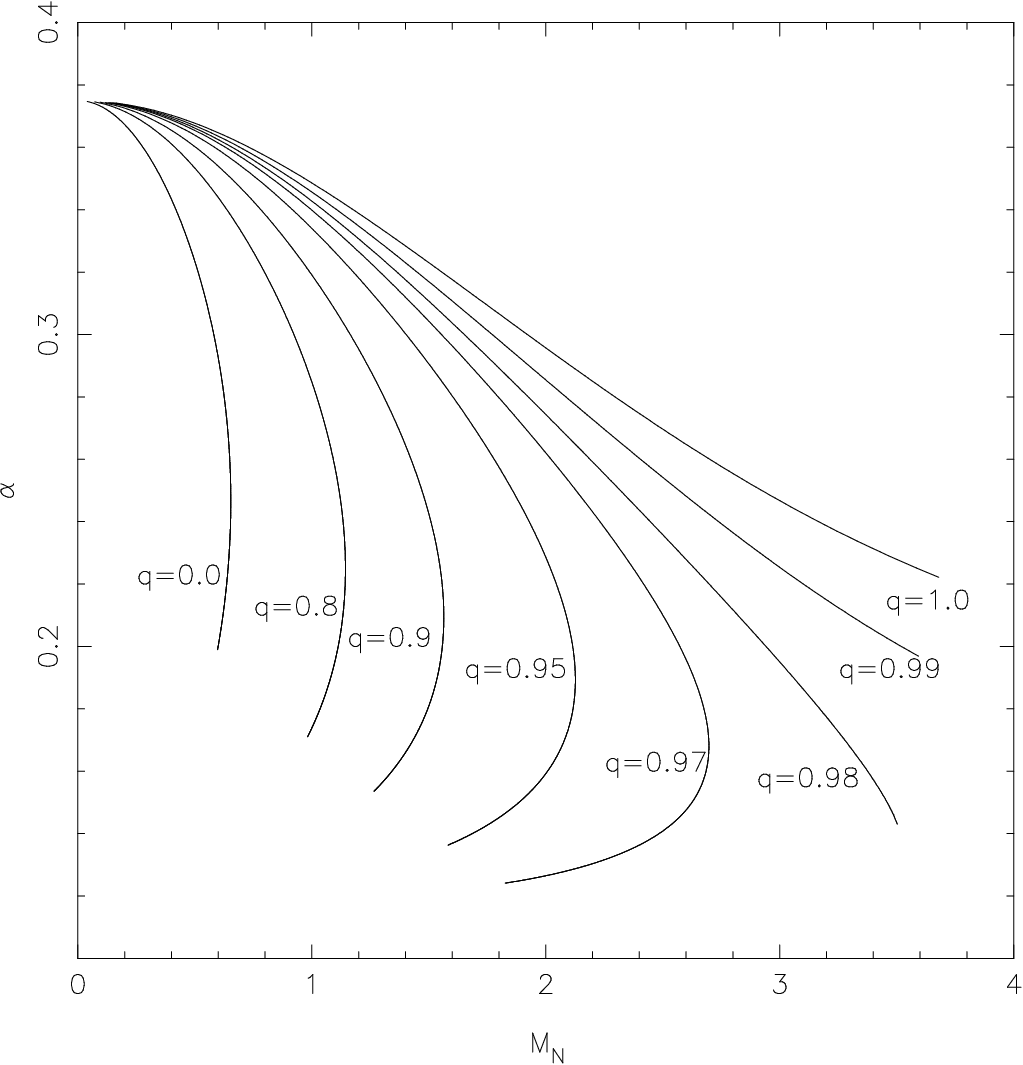}\\ 
\caption{Scalar coupling $\alpha:=-S/M_{N}$ as a function of rest mass 
for charged boson stars in the $n=4$ Power Law theory. The limiting value
as $M_{N} \rightarrow 0$ may be obtained analytically using Eq.\
(\ref{wfalpha}).} 
\end{figure} 

\begin{figure}
\centering 
\leavevmode\epsfxsize=8cm \epsfbox{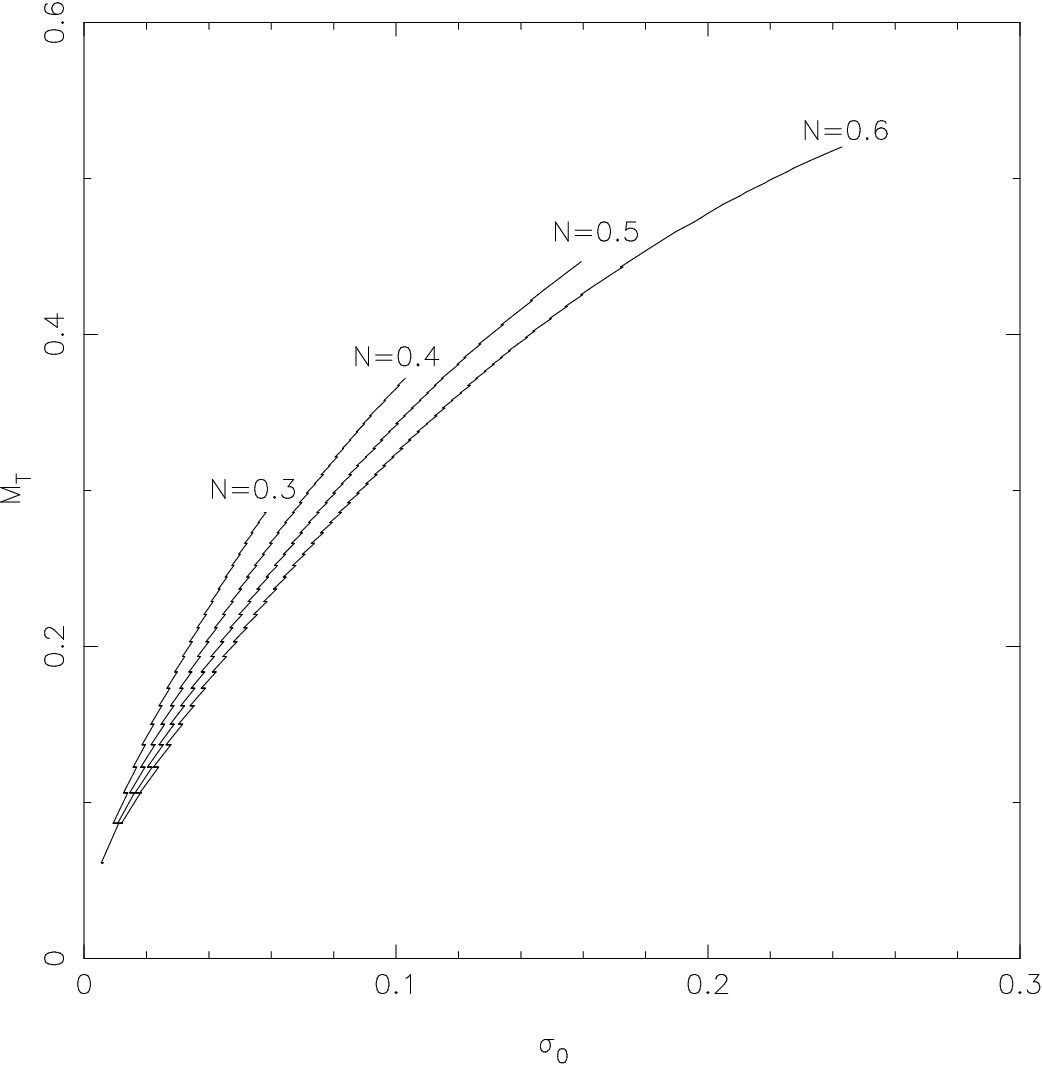}\\ 
\caption{Evolution curves for uncharged boson stars in $\omega=-1$
Brans Dicke theory with $\Lambda=0$. The curves are labelled by $N$
and evolve from configurations with $\Phi_{\infty}=1$ to 
configurations with $\Phi_{\infty}=6$. The direction of increasing 
time along the curves is towards the origin of the graph.}
\end{figure} 

\begin{figure}
\centering 
\leavevmode\epsfxsize=8cm \epsfbox{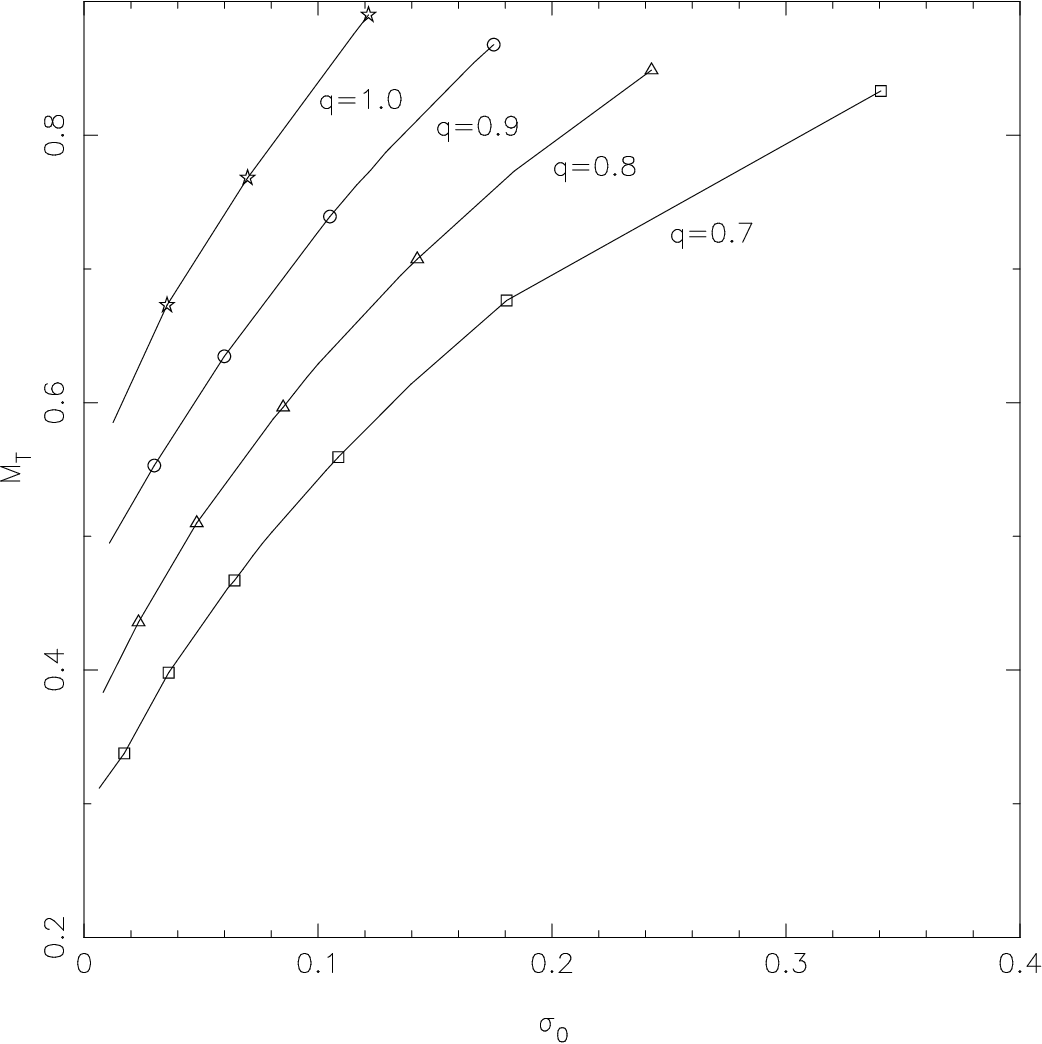}\\ 
\caption{Evolution curves for charged boson stars in $\omega=-1$
Brans Dicke theory with $\Lambda=0$. All stars have $N=1$ and evolve
{}from configurations with $\Phi_{\infty}=1$ to configurations with
$\Phi_{\infty}=2.8$. The direction of increasing time along the curves
is towards the origin of the graph. The symbols (squares, triangles, circles
and stars) on each curve are the
rescaled $\Phi_{\infty}=1$ generating solutions shown in Figure 2.}
\end{figure} 

\end{document}